\pdfoutput=1

\documentclass[num-refs]{wiley-article}

\input{preamble.sty}

\papertype{Submitted to Magnetic Resonance in Medicine}
\paperfield{MRM-18-19633}
\title{Fetal whole-heart 4D imaging using motion-corrected multi-planar real-time MRI}

\author[1]{Joshua FP van Amerom}
\author[1,2]{David FA Lloyd}
\author[1]{Maria Deprez}
\author[1]{Anthony N Price}
\author[1]{Shaihan J Malik}
\author[1,2]{Kuberan Pushparajah}
\author[1]{Milou PM van Poppel}
\author[1,3]{Mary A Rutherford}
\author[1,2]{Reza Razavi}
\author[1,3]{Joseph V Hajnal}

\affil[1]{School of Biomedical Engineering \& Imaging Sciences, King's College London, London, SE1 7EH, UK} 
\affil[2]{Department of Congenital Heart Disease, Evelina Children's Hospital, London, SE1 7EH, UK}
\affil[3]{Centre for the Developing Brain, King's College London, London, SE1 7EH, UK}

\corraddress{Joshua FP van Amerom, Department of Perinatal Imaging  \& Health, St Thomas' Hospital, London, SE1 7EH, UK}
\corremail{joshua.vanamerom@kcl.ac.uk}

\fundinginfo{Engineering and Physical Sciences Research Council~[EP/H046410/1]; Medical Research Council Strategic Fund [MR/K0006355/1]; Wellcome Trust [IEH Award 102431], the iFIND project; Wellcome/EPSRC Centre for Medical Engineering [WT203148/Z/16/Z]; National Institute of Health Research, Biomedical Research Centre, Guy’s and St Thomas’ National Health Services Foundation Trust and King’s College London}

\runningauthor{Fetal Whole-Heart 4D MRI \qquad van Amerom et al.}

\begin{document}

\maketitle

\begin{abstract}
\footnotesize{

\textbf{Purpose:} To develop a MRI acquisition and reconstruction framework for volumetric cine visualisation of the fetal heart and great vessels in the presence of maternal and fetal motion.

\noindent\textbf{Methods:} Four-dimensional depiction was achieved using a highly-accelerated multi-planar real-time balanced steady state free precession acquisition combined with retrospective image-domain techniques for motion correction, cardiac synchronisation and outlier rejection. 
The framework was evaluated and optimised using a numerical phantom, and evaluated in a study of 20 mid- to late-gestational age human fetal subjects. 
Reconstructed cine volumes were evaluated by experienced cardiologists and compared with matched ultrasound. 
A preliminary assessment of flow-sensitive reconstruction using the velocity information encoded in the phase of dynamic images is included.

\noindent\textbf{Results:} Reconstructed cine volumes could be visualised in any 2D plane without the need for highly-specific scan plane prescription prior to acquisition or for maternal breath hold to minimise motion. 
Reconstruction was fully automated aside from user-specified masks of the fetal heart and chest. 
The framework proved robust when applied to fetal data and simulations confirmed that spatial and temporal features could be reliably recovered.
Expert evaluation suggested the reconstructed volumes can be used for comprehensive assessment of the fetal heart, either as an adjunct to ultrasound or in combination with other MRI techniques.

\noindent\textbf{Conclusion:} The proposed methods show promise as a framework for motion-compensated 4D assessment of the fetal heart and great vessels.

}


\end{abstract}

\section{Introduction}  

Fetal cardiac MRI has been limited by the challenges associated with imaging a small, rapidly beating heart that is subject to various regular and spontaneous movements within the maternal torso\cite{Votino2012,Donofrio2014}.
However, as motion-tolerant fetal cardiac MR methods are introduced~\cite{Roy2017_MocoCineFetalCMR,vanAmerom2017} there is an increasing capacity for MRI to visualise the fetal heart and great vessels. 
The success of retrospective methods for high-resolution 3D depiction of the fetal brain\cite{Kuklisova-Murgasova2012,Gholipour2010} suggests the potential of this approach for 4D reconstruction of the fetal heart in the presence of motion, allowing for robust cardiac evaluation.
 
Recent published studies have taken advantage of the flexibility of continuous golden angle radial sampling and the resolution that can be achieved with segmented cine imaging using compressed sensing for 2D imaging of the fetal heart~\cite{Roy2017_CSMOG,Chaptinel2017,Haris2016FetalIGRASP}, with impressive results. 
In-plane motion-correction in k-space is possible with this approach~\cite{Roy2017_MocoCineFetalCMR}, however through-plane fetal and maternal movements cannot currently be corrected and in-plane motion across the entire field of view must be dealt with prior to or during cine MR image reconstruction.

This work builds on the motion-tolerant image-domain cine reconstruction framework previously described for 2D cardiac fetal imaging~\cite{vanAmerom2017} and methods for 3D image volume reconstruction~\cite{Kuklisova-Murgasova2012}, with the aim of developing an acquisition and reconstruction approach to generate a cine volume~(4D) representation of the fetal cardiovascular anatomy in utero from 2D multi-planar dynamic~(real-time) MR images, without the need for maternal breath-hold or significant manual processing.
Whole heart coverage presents challenges beyond 2D imaging, as motion must be corrected or compensated and the cardiac cycle synchronised for all acquired data is needed to allow for volumetric reconstruction.

In this work, 4D capability was achieved using a multi-planar real-time acquisition combined with retrospective image-domain techniques for motion correction, cardiac synchronisation and outlier rejection. 
Preliminary results showed the proposed framework was capable of reconstructing 4D cine volumes from dynamic MR images acquired without maternal breath-hold and without requiring precise scan plane prescription during acquisition~\cite{vanAmeromISMRM2018}.
In this work, the framework was validated using numerically simulated cardiac MRI data and evaluated in a study of twenty human fetal subjects. 
A preliminary assessment of velocity-sensitive reconstruction was also performed.

\section{Methods}  

\begin{SCfigure}[][htb]
\centering
\includegraphics[width = 0.46\textwidth]{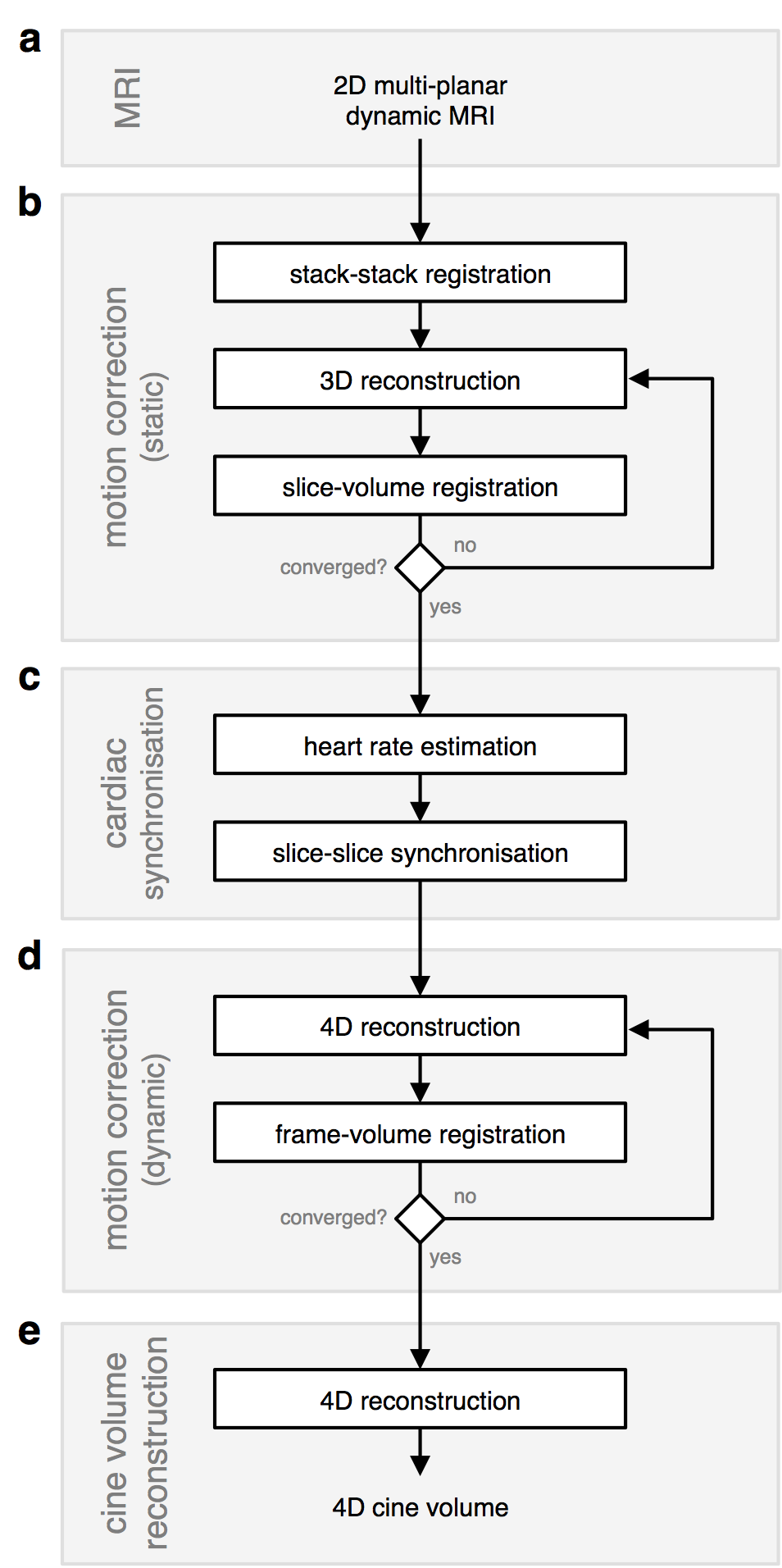}
\caption{Framework for 4D cine volume reconstruction, consisting of (\textbf{a})~acquisition and reconstruction of multi-planar dynamic 2D MRI;
(\textbf{b})~an initial motion correction stage to achieve rough spatial alignment of the fetal heart using temporal mean (i.e.,~static) images for stack-stack registration followed by slice-volume registration interleaved with static volume~(3D) reconstruction;
(\textbf{c})~cardiac synchronisation, including heart rate estimation and slice-slice cardiac cycle alignment; and 
(\textbf{d})~further motion-correction using dynamic image frames interleaved with 4D  reconstruction; and
(\textbf{e})~a final 4D cine reconstruction, including outlier rejection.
User-specified ROIs, and identification of a target stack for stack-stack registration are the only manual preparations required for reconstruction.}
\label{fig:cine_volume_reconstruction}
\end{SCfigure}

The proposed framework for 4D whole-heart reconstruction is shown in Figure~\ref{fig:cine_volume_reconstruction} and consists of motion correction using 'static' temporal mean images to achieve rough spatial alignment, followed by cardiac synchronisation and further motion correction using dynamic images, and concluding with a final 4D reconstruction including outlier rejection. 
The symbols used to denote the acquired dynamic MR images and reconstructed volume throughout this manuscript follow the convention commonly used for slice-to-volume reconstruction~(SVR) methods~\cite{Kuklisova-Murgasova2012,Gholipour2010}, which differs from those used previously to describe the purely 2D cine framework~\cite{vanAmerom2017}.

\subsection{Multi-Planar Dynamic MRI}  

Multi-planar data was acquired in stacks of parallel slices, as in previous SVR reconstruction studies\cite{Kuklisova-Murgasova2012}.
Stacks were acquired in multiple orientations to ensure full coverage of cardiac anatomy, and for each slice a series of real time images (or frames) was obtained, thus ensuring dense sampling in both space and time.
Fetal and maternal movement were expected to cause shifting of the fetal heart relative to the imaging field of view. 
These shifts can be large on the pixel scale and in any direction.

The acquired data form a set of $N_k$ dynamic MR image frames, $\mathbf{Y} = \left\{\mathrm{Y}_k\right\}_{k=1,\dots,N_k}$, where each frame has an associated acquisition time $t_k$ and consists of elements~$y_{jk}$ at 2D spatial coordinates indexed by~$j$. 
Subsets of $\mathbf{Y}$ for a single slice are defined as $\mathbf{Y}_{l} = \left\{\mathrm{Y}_k\right\}_{k \in \text{slice}_l}$ where parallel $\mathbf{Y}_{l}$ form a stack.

In the previous 2D+time framework~\cite{vanAmerom2017} dynamic MR images were reconstructed using k-t SENSE~\cite{Tsao2003} with spatially-adaptive regularisation to preserve the full temporal resolution of the accelerated acquisition. 
These complex-valued dynamic images were then combined as cine image series using robust statistics based on complex-valued errors to suppress voxels corrupted by artefact.
This was possible because the phase of the complex-valued dynamic images was fairly consistent in a single slice.
However, in this 4D framework, the phase of the dynamic images is inconsistent due to changes in slice orientation and position. 
As a result, these complex-valued images cannot be directly combined as a coherent complex-valued volume and magnitude-valued images were used. 
Consequently, magnitude-valued images from a k-t SENSE reconstruction were used to simplify the volume reconstruction methods and focus on the key challenge of locating slices in 3D space and time to reconstruct cine volumetric representations of the fetal cardiovascular system.

Motion correction, cardiac synchronisation and volume reconstruction were all performed on a region of interest~(ROI) centred on the fetal heart. 
In this work static ROIs covering the heart, great vessels and arterial arches were manually prescribed for each slice~$l$. 
A fetal chest volume of interest and target stack used for motion correction were also user-specified.
These were the only manual preparations required.

\subsection{Volume Reconstruction}  

Reconstruction of a 4D cine volume from 2D MR images requires information about the spatial and temporal position of the heart in each acquired 2D image frame. 
While the slice position and acquisition times are known, the relative spatial displacement due to fetal-maternal motion and the cardiac phase are not. 
However, if the spatial and temporal positions can be accurately estimated, the spatial location of the image frames can be transformed using rigid transformation matrices $\mathbf{A} = \left\{\mathrm{A}_k\right\}_{k=1,\dots,N_k}$, assuming rigid-body displacement, and acquisition times can be mapped to cardiac phases, $\boldsymbol{\theta} = \left\{\mathrm{\theta}_k\right\}_{k=1,\dots,N_k}$, with $\theta_k$ defined on a cyclic interval $[0,2\pi]$.

Image domain volume reconstruction is based on forward modelling of the image acquisition process, with volume reconstruction formulated as the solution to an inverse problem. 
In this work the aim was to recover a volumetric representation the beating fetal heart. 
The MR image acquisition model previously used for static anatomy~\cite{Kuklisova-Murgasova2012} was modified to include a temporal component, assuming the periodic motion of the cardiac anatomy can be characterised by a cardiac phase so that all acquired dynamic images can be combined as a single cardiac cycle. 
This MR image acquisition model describes the relationship between MR image frames, $\mathrm{Y}_k$, and a high resolution cine volume, $\mathbf{X} = \left\{\mathrm{X}_h\right\}_{h=1,\dots,N_h}$, with three spatial dimensions and a fourth periodic temporal dimension, 
\begin{equation} \label{eq:mr_image_aquisition_model}
\mathrm{Y}_k = \sum_h^{} \mathrm{W}^{hk} \mathrm{X}_h 
\end{equation}
where $\mathrm{X}_h$ has elements $x_{ih}$ for spatial index $i$ and temporal index $h$ corresponding to cardiac phases $\boldsymbol{\vartheta} = \left\{\vartheta_h\right\}_{h=1,\dots,N_h}$, and matrices $\mathrm{W}^{hk}$ are the product of a spatial weight, $\mathbf{M}^{k}$, and a temporal weight, $d^{hk}$, so that $w^{hk}_{ij} = d^{hk} m^{k}_{ij}$.
Each row of $\mathbf{M}^{k}$ is made of coefficients $\{m^{k}_{ij}\}_{i=1,\dots,N_i}$ that relate the spatial locations of cine volume $\mathbf{X}$ and MR image frame $\mathrm{Y}_k$, taking in to account spatial blurring and down-sampling, as well as the movement defined by spatial transformation $\mathrm{A}_k$. 
A point spread function~(PSF), determined by the MR acquisition, forms the basis for $\mathrm{W}^{hk}$.
The spatial in-plane and through-plane PSF were approximated as Gaussian~\cite{Kuklisova-Murgasova2012,Jiang2007}, while the temporal PSF for the band-limited k-t SENSE reconstruction was a sinc~\cite{vanAmerom2017}.

An initial estimate of $\mathbf{X}$ was obtained by PSF-weighted interpolation of the scattered data~\cite{Kuklisova-Murgasova2012,Rousseau2006}.
The error between acquired image frames and the image frames predicted by the acquisition model~(Eq.~\ref{eq:mr_image_aquisition_model}), given an estimate of $\mathbf{X}$, was calculated as 
\begin{equation} \label{eq:voxel_error}
e_{jk} = y^*_{jk} - \sum_j^{} \sum_k^{} w_{ij}^{hk} x_{ij}^{} \; ,
\end{equation}
where $y^*_{jk}$ are the elements of acquired image frames intensity-matched to account for variation in signal scaling and differential bias fields~\cite{Kuklisova-Murgasova2012}. 
Robust statistics were employed on both a voxel- and image frame-wise basis to reduce the impact of data corrupted by artefact or being misplaced in space or time.
The probabilities of a voxel, $p^\text{voxel}_{jk}$, or image frame, $p^\text{frame}_k$, being an inlier rather than outlier were combined as $p_{jk} = p^\text{voxel}_{jk} p^\text{frame}_k$.

PSF-weighted interpolation leads to blurring in the reconstructed volume due to the thick slices of the acquired images.
However, super-resolution~(SR) methods~\cite{Kuklisova-Murgasova2012,Gholipour2010} can be used to recover a high-resolution volume.
This was accomplished using gradient descent to minimise the error in Eq.~2 by solving 
\begin{equation} \label{eq:superresolution_minimistion_problem}
\mathbf{X} = \underset{\mathbf{X}}{\operatorname{argmin}} \left( \sum_j^{} \sum_k^{} p_{jk} \left\lVert e_{jk} \right\rVert ^2_2 + \lambda R(\mathbf{X}) \right) \;,
\end{equation}
which includes a spatial edge-preserving regularisation term, $R(\mathbf{X})$, to stabilise the reconstruction and a regularisation controlling parameter, $\lambda$. 
Intensity matching, bias correction, robust statistics and  $\mathbf{X}$ were updated at each  iteration~\cite{Kuklisova-Murgasova2012}.

\subsection{Motion Correction}  

To achieve the spatial coherence required for volume reconstruction, rigid body image registration was used to estimate transformations~$\mathbf{A}$ that align dynamic image frames~$\mathbf{Y}$ with cine volume~$\mathbf{X}$.
Registration was performed in three stages to facilitate convergence, as depicted in Figure~\ref{fig:cine_volume_reconstruction}.
The temporal mean of all dynamic images in a slice, $\overline{\mathrm{Y}}_l$, was used as a static reference free from cardiac pulsation. 
Use of static images for the initial stages of motion correction allowed for gross spatial alignment, facilitating cardiac synchronisation between slices and providing an initial estimate of $\mathbf{A}$ prior to motion correction of individual image frames.

Stacks were first aligned by volumetric registration with a target stack as in other volume reconstruction work~\cite{Kuklisova-Murgasova2012,Jiang2007,Rousseau2006}, using static temporal mean images $\overline{\mathbf{Y}}$.

Using the transformations estimated by stack-stack registration, an initial static volume was reconstructed from  $\overline{\mathbf{Y}}$ and then each slice $\mathrm{Y}_l$ was registered to the volume.
Interleaved volume reconstruction and slice-volume registration was repeated over multiple iterations to establish slice-wise alignment.

Frame-wise spatial alignment was performed following cardiac synchronisation.
An initial cine volume was reconstructed using estimated cardiac phases,~$\boldsymbol{\theta}$, and slice-wise transformations,~$\mathbf{A}$.
Subsequently, frame-volume registration was performed between each dynamic image frame, $\mathrm{Y}_k$ and the cine volume frame, $\mathrm{X}_h$, with matched cardiac phase, i.e.,~$\theta_h = \vartheta_k$, followed by cine volume reconstruction using the estimated frame-wise transformations.
Interleaved 4D reconstruction and frame-volume registration was repeated over multiple iterations.

Displacement was used to assess the change in position of the voxels in $\mathbf{Y}$ subject to estimated transformations $\mathbf{A}$.
Global displacement, $\text{disp}(\mathbf{A})$, was calculated as 
\begin{equation} \label{eq:transformation_displacement}
\text{disp}(\mathbf{A}) = \frac{\sum_{k}\sum_{j}^{} \text{dist}\left(y_{jk},\mathrm{A}_k(y_{jk})\right)}{\sum_{k} N_{j}} \; ,
\end{equation}
where $\text{dist}\left(y_{jk},\mathrm{A}_k(y_{jk})\right)$ is the spatial distance between the original position of voxel $y_{jk}$ and the position of $y_{jk}$ transformed by $\mathrm{A}_k$.
Similarly, deviation from the average slice transformation was used to assess the dispersion of estimated transformations.
Global deviation, $\text{dev}(\mathbf{A})$, was calculated as 
\begin{equation} \label{eq:transformation_deviation}
\text{dev}(\mathbf{A}) = \frac{\sum_{k}\sum_{j}^{} \text{dist}\left(\mathrm{A}_l(y_{jk}),\mathrm{A}_k(y_{jk})\right)}{\sum_{k} N_{j}} \; ,
\end{equation}
where the average slice transformation, $\mathrm{A}_l$, is the $p^\text{frame}$-weighted Frech\'{e}t mean~\cite{Aljabar2008}. 
Slice-wise deviation, $\text{dev}(\mathbf{A}_l)$, was calculated from Eq.~\ref{eq:transformation_deviation} by restricting the summations to image frames $k$ acquired at slice location $l$.

\subsection{Cardiac Synchronisation}  

\begin{SCfigure}[][htb]
\centering
\includegraphics[width = 0.48\textwidth]{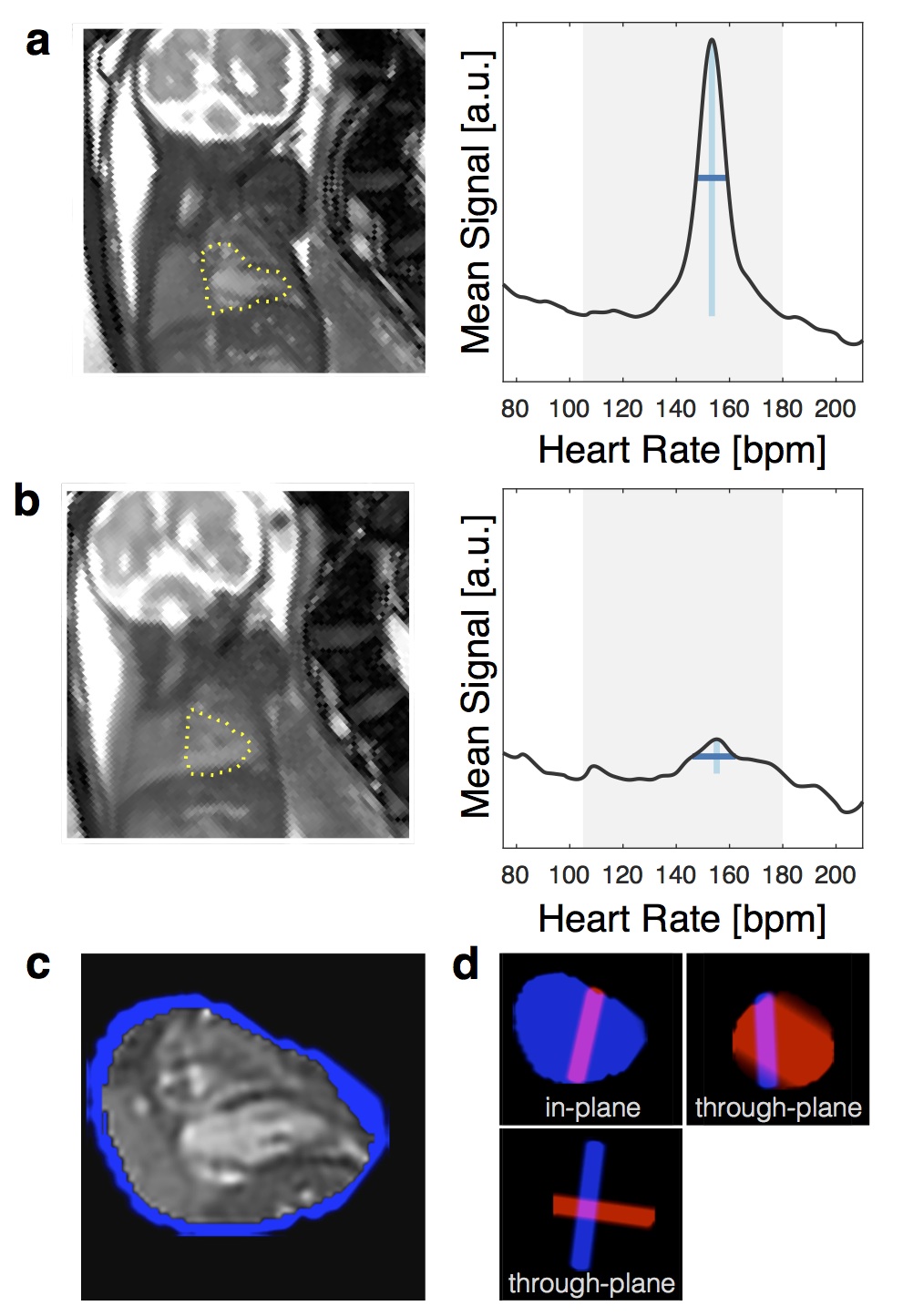}
\caption[Cardiac Synchronisation]{Cardiac synchronisation in healthy 30\textsuperscript{+3}~week gestational age fetus. 
Heart rate estimation: (\textbf{a})~Temporal mean image, $\overline{\mathrm{Y}}_l$, for a slice with little motion~(${\text{dev}(\mathbf{A}_l) = \text{1.5 mm}}$) with ROI over the fetal heart~(dotted yellow line), rotated to show the fetus in radiographic orientation, and plot of temporal frequency spectrum~(spatial mean in the fetal heart ROI of the temporal Fourier transform of the dynamic image series) showing a clear maxima in the range of expected heart rates (grey band, 110--180~bpm). Peak height~(light blue vertical line) and width~(dark blue horizontal line) are shown.
(\textbf{b})~Fetal movement during the acquisition of a different slice from the same stack~(${\text{dev}(\mathbf{A}_{l}) = \text{9.4 mm}}$) leads to blurring in the temporal mean image, as well as a lower and  broader peak in the temporal frequency spectrum. Unreliable heart rates, such as (b), were interpolated from temporally adjacent slices in the same stack.
Slice-slice synchronisation: (\textbf{c})~Single frame of the cine, $\mathbf{X}_l$, corresponding to slice shown in (a) and (\textbf{d})~volume weights, $\mathrm{V}_{l}$,~(blue region), as in-plane and orthogonal through-plane views showing intersection with a slice from a different stack~(red region).
Slice-slice cardiac synchronisation was performed by temporally aligning the cardiac cycles between $\mathbf{X}_l$ constructed from $\mathbf{Y}_l$ while using volume weights, $\mathbf{V}_l$, to determine the overlap between slices.}
\label{fig:method_cardiac_synchronisation}
\end{SCfigure}

The cardiac phase of each image frame must be known for the data to be combined as a single cardiac cycle.
The heart rate was first estimated independently for each slice and then the cardiac cycle was synchronised between slices, as shown in Figure~\ref{fig:method_cardiac_synchronisation}.

A heart rate was estimated for each slice from the temporal frequency spectrum, i.e., the temporal Fourier transform of the dynamic image series, by identifying the maxima in the spatial mean of the temporal frequency spectrum in the fetal heart ROI~\cite{vanAmerom2017} (Fig.~\ref{fig:method_cardiac_synchronisation}a).
Displacement of the heart and variation of the fetal heart rate resulted in diminished peaks in the temporal frequency spectra~(Fig.~\ref{fig:method_cardiac_synchronisation}b). 
Unreliable heart rate estimates were identified as those with peak height or width more than three normal standard deviations from the median, estimated from the median absolute deviation~\cite{Rousseeuw1993}. 
Excluded estimates were replaced by the linear interpolation of the heart rates in temporally adjacent slices within the same stack.

The cardiac cycle was synchronised across slices by estimating a temporal offset for each slice that aligns the pulsation of the heart in overlapping slices.
To determine the overlap between slices in 3D space, dynamic MR images were cast as cines in volume space.
Specifically, a cine volume,~$\mathbf{X}_{l}$, was reconstructed for each slice,~$l$, from the dynamic images in that slice,~$\mathbf{Y}_l$.
Volume weights,~$\mathbf{V}_{l}$, representing the weighted contribution of the image frames acquired at slice location~$l$ in volume space, ${v_{ihl} = { \sum_j{ \sum_{k \in \text{slice}_l}{ w^{hk}_{ij} } } }}$, were used to determine the spatial intersection of slices.
The overlap between slice~$l$ and slice~$l'$~(Fig.~\ref{fig:method_cardiac_synchronisation}d) was calculated as $\sum_{i}^{}\sum_{h}^{}v_{ihl}v_{ihl'}$.

Applying a temporal offset, $\theta^\text{offset}_l$, to $\mathbf{X}_{l}$ was achieved by applying a cyclic Fourier time shift, $\varphi_{\theta^\text{offset}}(\:)$, i.e.,~addition of a linear phase in the temporal frequency domain.
Aligning the cardiac cycles of all the slices was equivalent to determining $\{\theta^\text{offset}_l\}_{l=1 \dots N_l}$ that maximised the overlap-weighted similarity between all $\varphi_{\theta^\text{offset}_l}\left(\mathbf{X}_{l}\right)$. 
A group-wise optimisation was computationally intensive in practice, so a bootstrap approach was adopted instead, where one temporal offset was estimated at each iteration, i.e.,
\begin{equation}
\underset{\theta^\text{offset}_{l}}{\operatorname{argmin}} \quad -\sum_{l' \in \{ l''\}}^{} \frac{ \sum_{i}^{}\sum_{h}^{}v_{ihl}v_{ihl'}\;\rho\left( \varphi_{\theta^\text{offset}_l}\left(\mathbf{X}_{l}\right),\varphi_{\theta^\text{offset}_{l'}}\left(\mathbf{X}_{l'}\right) \right) }{ \sum_{i}^{}\sum_{h}^{}v_{ihl}v_{ihl'} } \; ,
\label{eq:slice_slice_synchronisation}
\end{equation}
where similarity measure $\rho(\:)$ is Pearson's correlation and \{ l''\} is a set of target slices.
The slice $l'$ with the greatest overlap with all other slices was assigned ${\theta^\text{offset}_{l'} = 0}$, and used as an initial target slice.
A different slice with the greatest overlap with $l'$ was then identified and $\theta_l^\text{offset}$ was estimated by Eq.~\ref{eq:slice_slice_synchronisation} and cardiac phases were adjusted as ${\theta_k = (\theta_k+\theta^\text{offset}_l ) \; \textit{mod} \; 2\pi}$.
This slice was then added to the set of fixed target slices~$\{l''\}$ and the process was repeated using the next slice with the greatest overlap with $\{l''\}$. 
This continued until all $\theta^\text{offset}_{l}$ were estimated.

\subsection{Fetal Study}  

Multi-planar dynamic MR imaging was acquired in a cohort of eleven consecutive singleton pregnancies as part of a larger research project when examination time permitted acquisition of a minimum of three stacks with a combined total of at least thirty slices to ensure good coverage of cardiovascular anatomy. 
The cases in this initial cohort were used for method development and evaluation. 
Subsequently, data was acquired in a second cohort of nine fetuses, eight singleton and one twin pregnancy~(ID13), at the discretion of the attending cardiologist without a physicist present at the examination to facilitate robust stack prescription.
These cases were used to assess the utility of the methods.
Fetal subjects ranged from 23 to 33 weeks gestational age.
The study was conducted with the approval of the local research ethics committee and all participants gave written informed consent prior to enrolment.
Details of all fetal cases are listed in Table \ref{tab:fetal_subjects}, comprising three healthy cases and seventeen with cardiac anomalies.

\begin{table}[htbp]
\caption[Fetal Study Subjects]{Fetal study subjects.}
\centering
\begin{tabular}{ c c l c c c }
\headrow
\textbf{ID} & \textbf{GA} & \textbf{Reason for Scan} & \textbf{No.\ Stacks} & \textbf{No.\ Slices} & \textbf{US} \\
\multicolumn{6}{l}{\textbf{Cohort 1}} \\
\hline 
01 & 32\textsuperscript{+1} & dilated aortic root & 5 & 37 & \textasteriskcentered \\
02 & 30\textsuperscript{+3} & volunteer & 5 & 37 & \textasteriskcentered \\
03 & 24\textsuperscript{+6} & volunteer & 5 & 45 & \textasteriskcentered \\
04 & 29\textsuperscript{+6} & right aortic arch &  6 & 54 & \\     
05 & 31\textsuperscript{+0} & ventricular diverticulum &  6 & 55 & \textasteriskcentered \\
06 & 32\textsuperscript{+3} & right aortic arch &  6 & 75 & \\
07 & 31\textsuperscript{+4} & double aortic arch &  4 & 44 &  \\
08 & 32\textsuperscript{+2} & right aortic arch &  4 & 42 & \textasteriskcentered \\
09 & 28\textsuperscript{+0} & volunteer &  5 & 51 & \textasteriskcentered \\
10 & 31\textsuperscript{+4} & double aortic arch &  3 & 35 & \textasteriskcentered \\
11 & 33\textsuperscript{+2} & cardiac tumour &  4 & 42 & \\
\hline 
\multicolumn{6}{l}{\textbf{Cohort 2}} \\
\hline 
12 & 32\textsuperscript{+0} & interrupted aortic arch & 3 & 25 & \\
13 & 30\textsuperscript{+5} & cardiac tumour & 3 & 39 & \\
14 & 31\textsuperscript{+3} & common arterial trunk & 3 & 33 & \\
15 & 30\textsuperscript{+4} & transposition of the great arteries & 4 & 45 & \\
16 & 31\textsuperscript{+6} & right aortic arch & 4 & 23 & \\
17 & 31\textsuperscript{+4} & dextrocardia & 3 & 34 & \\
18 & 23\textsuperscript{+5} & partial anomalous pulmonary venous return & 3 & 33 & \\
19 & 32\textsuperscript{+1} & dilated ascending aorta & 3 & 33 & \\
20 & 32\textsuperscript{+4} & right aortic arch & 3 & 33 & \\
\hline 
\end{tabular}
\begin{tablenotes}
\item \textbf{ID}, fetal case number; \textbf{GA},  gestational age in weeks\textsuperscript{+days}; \textbf{No.\ Stacks}, number of acquired multi-planar dynamic MRI stacks; \textbf{No.\ Slices}, number of slices acquired across all stacks; \textbf{US}, matched 2D and STIC ultrasound data acquired within three days of MRI examination; \textbf{Cohort 1}, eleven consecutive singletons scanned for method development and evaluation; \textbf{Cohort 2}, eight singletons and one twin scanned for clinical assessment. 
\end{tablenotes}
\label{tab:fetal_subjects}
\end{table}

\subsubsection{Imaging}  

Stacks of multi-planar dynamic images were acquired on a 1.5~T Ingenia MR~system~(Philips, Netherlands) using a 2D balanced steady state free precession~(bSSFP) sequence. 
Highly-precise scan plane prescription was not required, as specific views could be later obtained from the reconstructed 4D data. 
However, to achieve a good distribution of scan plane orientations, the first three stacks were prescribed roughly transverse, sagittal and coronal to the fetal trunk for all cases in cohort~1, with additional stacks prescribed in scanner transverse, sagittal, or coronal planes.
In-plane rotation was used to reduce the size of the field of view in the phase encode direction while avoiding alignment of pulsatile maternal anatomy (e.g.,~maternal descending aorta) with the fetal heart in the phase encode direction to avoid complications in the k-t SENSE reconstruction.

Sequence parameters were selected to balance competing goals of good signal and high spatio-temporal resolution with full coverage of the fetomaternal anatomy in the field of view, minimal scan time and safety constraints. 
Scanner operation was constrained to ensure fetal and maternal safety with respect to specific absorption rate~(whole body SAR~<~2.0~W/kg~\cite{Hand2010}), peripheral nerve stimulation~(low PNS mode), and sound pressure level~(SPL~<~85~dB(A), accounting for >30~dB attenuation in utero~~\cite{Glover1995}).
While necessary, safety constraints placed a limit on scanner performance and, consequently, achievable resolution and image quality.

Multi-planar bSSFP data were collected with regular Cartesian k-t undersampling~\cite{Tsao2005} using the following default parameters: 
TR/TE~3.8/1.9~ms, 
flip angle~60\textdegree, 
FOV~400×304~mm, 
voxel size~2.0×2.0×6.0~mm, 
8×~acceleration, 
72~ms~temporal resolution,
96~images per slice, 
slice overlap~2--3~mm. 
A steady state was established using an ${\alpha\text{/2-TR/2}}$~preparation pulse~\cite{Deimling1994} and dummy excitations.
Coil calibration data were acquired in a scan immediately preceding the multi-planar acquisition, and low spatial-resolution training data was acquired immediately following the under-sampled data.
Acquisition of a typical stack took 155~s. 
In some cases the default FOV was either decreased or increased in the phase encode direction to accommodate maternal anatomy, with a proportional change in the temporal resolution~(cohort~1 median decrease 7.6~ms, median increase 3.8~ms).

Ultrasound~(US) examinations were performed in seven cases in cohort~1, as indicated in Table~\ref{tab:fetal_subjects}, using an EPIQ V7G system with C9-2 curved array, V6-2 curved volume and X6-1 matrix array transducers~(Philips, Netherlands).
The echocardiography protocol included comprehensive 2D M-mode imaging as well as B-mode sweeps covering the fetal trunk that were used to produce 4D Spatio-Temporal Image Correlation~(STIC) volumes~\cite{DeVore2003}.

\subsubsection{Reconstruction}  

All k-t SENSE reconstructions were performed in MATLAB~(Mathworks, USA), with additional functionality from  ReconFrame~3.0.535~(GyroTools, Switzerland) to place the data in a common 3D space. 
Slice-slice synchronisation~(Eq.~\ref{eq:slice_slice_synchronisation}) was achieved by constrained non-linear multivariate minimisation using the interior point approach implemented in MATLAB's Optimisation Toolbox.

Volume reconstruction was performed using the Image Registration Toolkit~(BioMedIA, UK), building on the framework implemented by Kuklisova \textit{et al.}~\cite{Kuklisova-Murgasova2012} with additional functionality for dynamic data.
Reconstruction parameters were based on those previously described~\cite{Kuklisova-Murgasova2012}, and validated for this work using simulation experiments. 
Volume reconstruction was performed at a spatial resolution of 1.25~mm, or 5/8 of the actual acquired in-plane resolution, similar to the ratio used for reconstruction of fetal brain MRI~\cite{Kuklisova-Murgasova2012}, and a temporal resolution corresponding to $N_{h}=\text{25}$ cardiac phases, to provide frame-volume registration targets across the entire cardiac cycle.
A user-specified volume of interest covering the fetal chest was used for slice-volume registration, making use of the anatomy surrounding the heart to facilitate the registration process, while user-specified 2D fetal heart ROIs were combined as a volume and used for frame-volume registration.
Reconstruction code can be found at \href{https://github.com/jfpva/fetal\_cmr\_4d}{github.com/mriphysics/fetal\_cmr\_4d}~(SHA1:b6c94f073c).

\subsubsection{Evaluation}  

Whole-heart 4D cine volumes were assessed by expert cardiac MRI readers to establish the quality of the reconstruction method.
Three observers~(DL, KP, MvP) with four, seven, and one years experience reading cardiac MRI, respectively, were asked to independently navigate the reconstructed volumes using the Medical Image Interaction Toolkit Workbench~(DKFZ, Germany) and score them in eleven categories, based on the segmental approach to defining cardiac anatomy and pathology~\cite{Donofrio2014,Yoo1999,Carvalho2005}.
Data was presented to the reviewers without context, though the reviewers may have been present at the time of scanning and may have recalled unique cases.
A five-point scale was used for scoring, as follows:
\begin{enumerate}[nosep]
\item[] 4: high image quality and distinct appearance of cardiac structures;
\item[] 3: adequate image quality to determine most details;
\item[] 2: sufficient image quality to determine some details;
\item[] 1: poor image quality with significant lack of detail; and
\item[] 0: inadequate image quality to visualize any cardiac structure.
\end{enumerate}
Ultrasound images acquired during the broader research project were used for qualitative comparison with 4D cine volumes in cases where matched data were available.

\subsection{Simulation}  

\begin{figure}[htb]
\centering
\includegraphics[width = 0.71\textwidth]{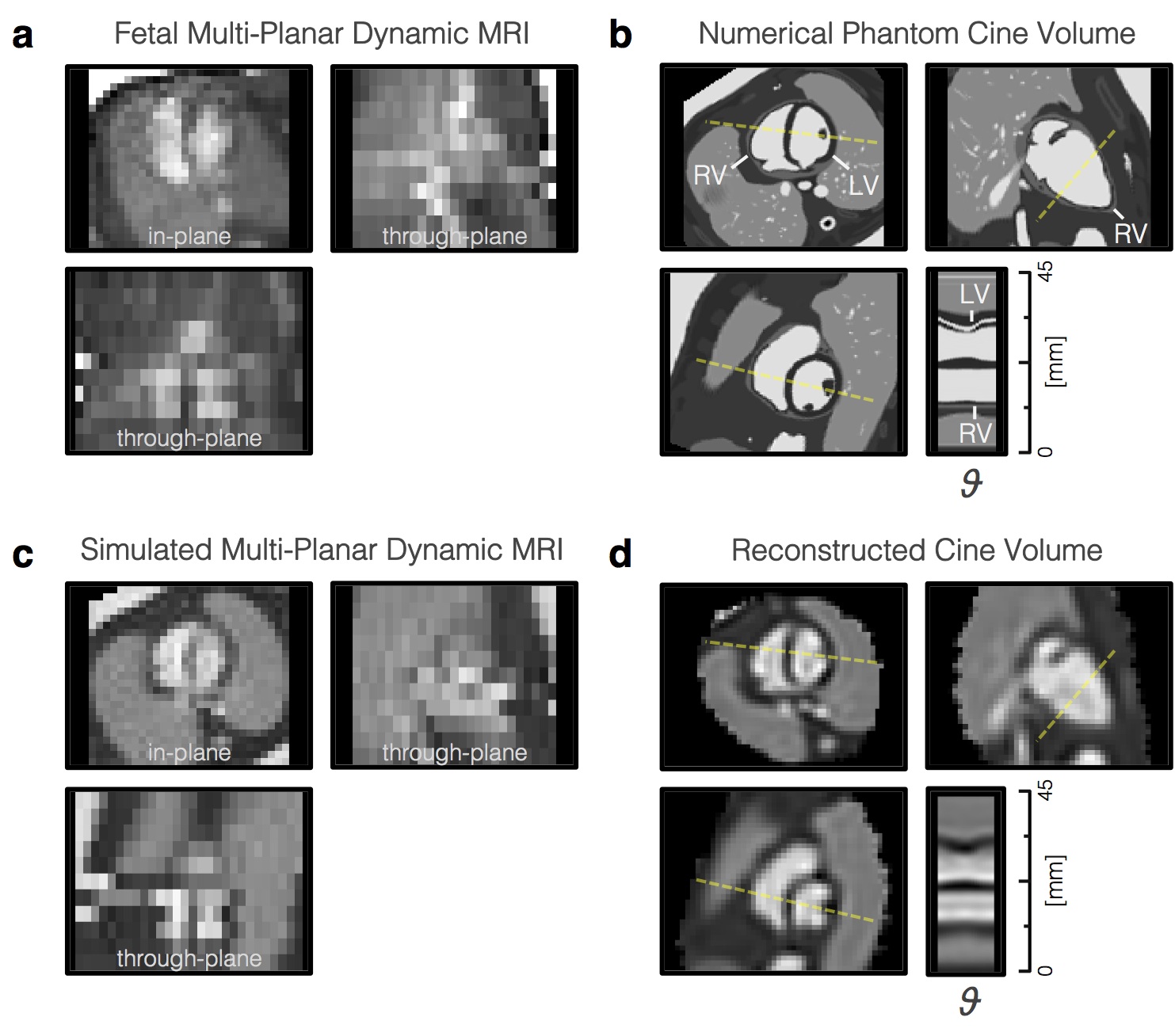}
\caption[Simulation Data Used to Optimise and Evaluate Proposed Methods]{Simulation data used to optimise and evaluate the proposed methods. 
(\textbf{a})~Dynamic MR image frame of the heart in an early third trimester fetus~(top left), acquired during the fetal study, showing misalignment of slices due to fetal movement in the through-plane views~(bottom and right). 
(\textbf{b})~High spatio-temporal resolution cine volume generated from numerical phantom of beating fetal heart, shown in orthogonal planes as in (a), with line profile~(yellow dashed line) showing contraction and dilation of left~(LV) and right~(RV) ventricles across the phases~($\vartheta$) of the cardiac cycle. 
(\textbf{c})~Simulated multi-planar dynamic MR images derived from numerical phantom in (b) with resolution, receiver noise and motion similar to the in utero MRI in~(a). 
(\textbf{d})~Cine volume reconstructed from simulated MRI data using the known cardiac phase and spatial transformation of each dynamic MR image frame.}
\label{fig:sim_mri}
\end{figure}

A numerically simulated phantom was used to optimise and evaluate the proposed methods under controlled conditions, as demonstrated in Figure~\ref{fig:sim_mri}. 
The MRXCAT phantom~\cite{Wissmann2014} was used to generate a cine volume scaled to the size of the fetal heart, $\boldsymbol{\chi}$, with high spatio-temporal resolution~(0.44~mm isotropic, 25~cardiac phases) and tissue properties adapted to the in utero environment, i.e., the signal of air in the lungs, trachea and the space surrounding the body in the postnatal simulation were replaced with the signal characteristics of amniotic fluid. 

Dynamic MR images with 0.5~mm in-plane resolution and 6~mm through-plane resolution were generated from $\boldsymbol{\chi}$ using Eq.~\ref{eq:mr_image_aquisition_model} and  
further down-sampled to 2~mm in-plane resolution by truncating k-space while adding noise.
Cardiac phases and displacements were based on in utero measurements of the fetal heart made during the fetal study. 
Transformation matrices were scaled to change the amount of displacement by a scalar factor~\cite{Alexa2002} to assess varying degrees of motion.
Cine volumes were reconstructed from simulated dynamic MR images and the results were compared with the ground truth data to evaluate the 4D reconstruction framework.

\subsection{Velocity-Sensitive Volume Reconstruction}  

Volume reconstruction from complex-valued dynamic images is not a straightforward extension of the proposed framework, but has the potential to generate cine velocity volumes, i.e., 4D velocity maps, from the same data.
Phase contrast bSSFP methods have been proposed~\cite{Markl2003}, using the inherent velocity-sensitivity of the bSSFP sequence to encode velocity in the phase images~\cite{Markl2004}. 
For complex-valued image frames, $\widetilde{\mathbf{Y}}$, the phase of an acquired voxel, $\angle \widetilde{y}_{jk}$, is the sum of a velocity-encoded phase, $\phi_{jk}$, and background phase.
The smoothly varying background phase can be estimated by fitting a low-order polynomial to the phase in static tissues and subtracted from $\angle \widetilde{y}_{jk}$ so that only $\phi_{jk}$ remains~\cite{Nielsen2009}.
The $\phi_{jk}$ can then be related to the underlying velocity as 
\begin{equation} \label{eq:velocity_encoded_phase}
\phi_{jk} = \gamma \left( \upsilon^\text{read}_{jk} M^\text{read}_{1} + \upsilon^\text{phase}_{jk} M^\text{phase}_{1} + \upsilon^\text{slice}_{jk} M^\text{slice}_{1} \right)
\end{equation}
where $\gamma$ is the gyromagnetic ratio, and $\upsilon$ are the velocities and $M_1$ are the first moments of the gradient waveforms in the readout, phase-encode and slice directions~\cite{Markl2003}.

To reconstruct volumetric velocity maps, the velocity-encoded phase would need to be related to three-dimensional space, and the image acquisition model~(Eq.~\ref{eq:mr_image_aquisition_model}) would have to be adjusted to include velocity terms, precipitating a number of changes to the framework. 
Instead, as a proof of principle, the framework was modified to reconstruct complex-valued $\widetilde{\mathbf{X}}$.
Background phase was removed from $\widetilde{\mathbf{Y}}$ by subtracting a third-order 3D polynomial fit to the phase in static amniotic fluid and tissue in each acquired stack and a phase sign correction was applied to improve phase consistency, so that the signs of all $\phi_{jk}$ were aligned with the target stack.
All steps in the framework were performed as described previously, with the modification that real- and imaginary-valued cine volumes were also generated separately from the real and imaginary components of $\widetilde{\mathbf{Y}}$.

\section{Results} 

Simulation experiments were used to evaluate the proposed method and select appropriate reconstruction parameters (Supporting Figure~\ref{fig:simulation_nrmse_v_delta_v_lambda}). 
The volume reconstructed from simulated MR images using the established parameters~(Fig.~\ref{fig:sim_mri}d) showed cardiac anatomy and pulsation matching the numerically simulated cine volume, providing evidence of geometric accuracy. 
However, there was some blurring in all reconstructed cine volumes compared to the high resolution numerical phantom~(Fig.~\ref{fig:sim_mri}b), due to the low spatio-temporal resolution sampling of the dynamic images, as well as signal inhomogeneity within the blood pool in the heart. 

Fetal results are summarised in Table~\ref{tab:fetal_results}, including expert scores as an indication of image quality.
Cine volumes were successfully reconstructed in all eleven fetal cases in the first cohort. 

\begin{table}[htbp]
\definecolor{lightgray}{gray}{0.9}
\caption[Summary of Fetal Results]{Summary of fetal results.}
\centering
\begin{tabular}{ c c c c c c c c c }
\headrow
\textbf{ID} & \textbf{Expert Score} & \multicolumn{2}{c}{\textbf{Heart Rate}} & \multicolumn{2}{c}{\textbf{Transformations}} & \multicolumn{2}{c}{\textbf{Outliers}} & \textbf{No. Frames} \\
\headrow
& [0--4] & \multicolumn{2}{c}{{\small [bpm]}} & \multicolumn{2}{c}{{\small [mm]}} & \multicolumn{2}{c}{{\small [\% $p<0.5$]}} &  \\
\headrow
 & & & & $\text{disp}(\mathbf{A})$ & $\text{dev}(\mathbf{A})$ & $p^\text{voxel}$ & $p^\text{frame}$ &  \\
\hline
\multicolumn{9}{l}{\textbf{Cohort 1}} \\
\hline
01 & 3.5 & \multicolumn{2}{l}{\; 147 $\pm$  5} &  4.6 & 1.1 & 0.4 &  5.7 & 3552 \\
02 & 3.6 & \multicolumn{2}{l}{\; 155 $\pm$  7} &  5.2 & 1.7 & 0.5 &  9.4 & 3552 \\
03 & 3.4 & \multicolumn{2}{l}{\; 147 $\pm$  6} &  7.7 & 1.3 & 1.1 &  7.4 & 3936 \\
04 & 3.1 & \multicolumn{2}{l}{\; 143 $\pm$  7} & 11.2 & 1.8 & 0.4 & 21.6 & 3163 \\
05 & 3.4 & \multicolumn{2}{l}{\; 148 $\pm$ 11} & 12.2 & 2.3 & 0.4 & 10.4 & 5280 \\
06 & 3.6 & \multicolumn{2}{l}{\; 139 $\pm$  4} &  7.6 & 1.3 & 0.5 & 15.6 & 6719 \\
07 & 3.1 & \multicolumn{2}{l}{\; 153 $\pm$  8} &  8.4 & 2.3 & 0.5 & 28.3 & 3072 \\
08 & 3.5 & \multicolumn{2}{l}{\; 151 $\pm$  5} &  4.9 & 1.2 & 0.4 & 20.2 & 3525 \\
09 & 3.4 & \multicolumn{2}{l}{\; 150 $\pm$  5} &  2.8 & 1.0 & 1.0 &  7.7 & 4887 \\
10 & 3.6 & \multicolumn{2}{l}{\; 150 $\pm$  3} &  4.7 & 1.3 & 0.8 &  4.6 & 3072 \\
11 & 3.2 & \multicolumn{2}{l}{\; 153 $\pm$  7} &  5.8 & 2.0 & 1.5 & 43.7 & 4032 \\
\hiderowcolors
\hline
\headrow
\rowcolor{white}
{\small Median} & 3.4 & \multicolumn{2}{l}{\; 150 $\pm$  6}  & 5.8 & 1.3 & 0.5 & 10.4 & 3552 \\
\hline
\rowcolor{white}
\multicolumn{9}{l}{\textbf{Cohort 2}} \\
\hline
\rowcolor{lightgray}
12 & 3.2 & \multicolumn{2}{l}{\; 145 $\pm$ 19} &  7.8 & 3.0 & 0.3 & 27.1 & 2304 \\
\rowcolor{white}
13 & 3.6 & \multicolumn{2}{l}{\; 125 $\pm$  1} &  4.7 & 0.7 & 0.7 & 13.0 & 3743 \\
\rowcolor{lightgray}
14 & 3.4 & \multicolumn{2}{l}{\; 148 $\pm$  7} &  9.0 & 1.7 & 0.6 & 18.4 & 2783 \\
\rowcolor{white}
15 & 3.6 & \multicolumn{2}{l}{\; 136 $\pm$  3} &  4.8 & 0.9 & 0.7 & 15.3 & 4121 \\
\rowcolor{lightgray}
16 & 2.5 & \multicolumn{2}{l}{\; 136 $\pm$  3} & 11.8 & 2.6 & 1.3 & 10.9 & 1920 \\
\rowcolor{white}
17 & 3.9 & \multicolumn{2}{l}{\; 127 $\pm$  4} &  6.0 & 1.0 & 0.5 & 17.3 & 3264 \\
\rowcolor{lightgray}
18 & 2.1 & \multicolumn{2}{l}{\; 142 $\pm$ 17} & 15.5 & 2.2 & 0.8 & 22.0 & 2590 \\
\rowcolor{white}
19 & 3.9 & \multicolumn{2}{l}{\; 141 $\pm$  4} &  5.4 & 0.5 & 2.0 & 12.1 & 3168 \\
\rowcolor{lightgray}
20 & 3.8 & \multicolumn{2}{l}{\; 137 $\pm$  2} &  3.8 & 0.9 & 0.9 & 11.5 & 3168 \\
\hline
\end{tabular}
\begin{tablenotes}
\item \textbf{ID}, fetal case number; \textbf{Expert Score}, mean of scores assigned across all categories by all reviewers~(Cohort~1) or reviewer~1 only~(Cohort~2); \textbf{Heart Rate}, mean and standard deviation of estimated fetal heart rates; \textbf{Transformations}, global displacement, $\text{disp}(\textsc{A})$, and deviation from average slice transformation, $\text{dev}(\mathbf{A})$; \textbf{Outliers}, percent of voxel,~$p^\text{voxel}$, and image frame,~$p^\text{frame}$, probabilities below 0.5; \textbf{No. Frames}, number of image frames contributing to volume reconstruction, prior to outlier rejection. Images at the edges of the volume, i.e., those contributing fewer than~$\textit{median}(N_j)/10$ voxels, were excluded from these summary statistics as they had minimal impact on the reconstructed volume but were more likely to be misaligned in space or time; \textbf{Cohort 1}, consecutive singletons scanned for method development and evaluation; \textbf{Cohort 2}, fetuses scanned for clinical assessment; \textbf{Median}, Cohort~1 median values.
\end{tablenotes}
\label{tab:fetal_results}
\end{table}

\begin{figure}[htb]
\centering
\bigskip
\includegraphics[width = 0.99\textwidth]{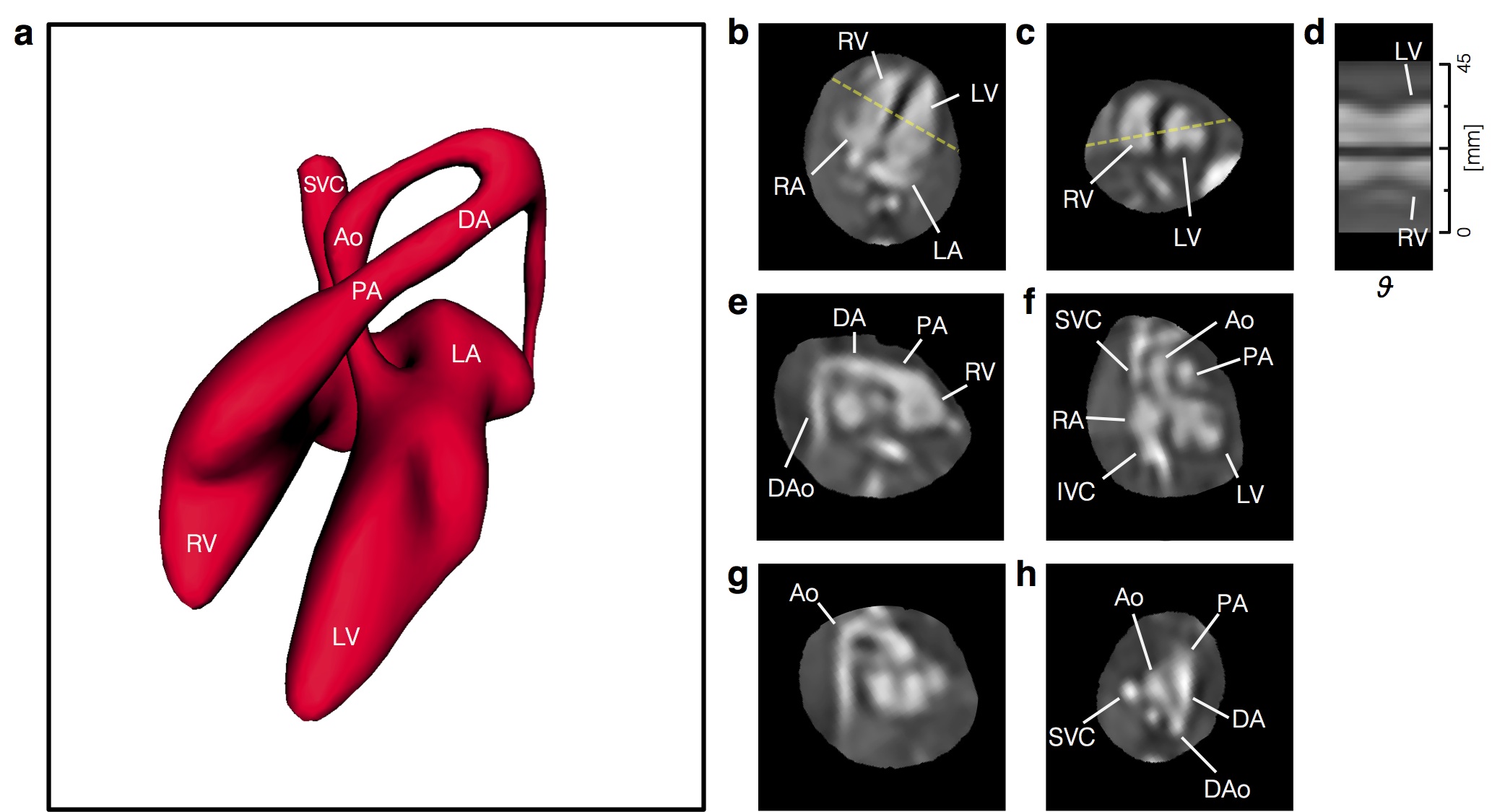}
\caption[Reconstructed 4D Cine MRI Volume of the Heart of a Healthy Fetus]{Reconstructed 4D cine volume of the heart of a healthy 28\textsuperscript{+0} week gestational age fetus (ID09). 
(\textbf{a})~Volume rendering of blood pool in diastole showing arrangement and connections of chambers and vessels, for reference. 
The reconstructed 4D data is shown re-sliced in  (\textbf{b})~four chamber, (\textbf{c})~mid-short axis, (\textbf{e})~right ventricular outflow tract, (\textbf{f})~left ventricular outflow tract, (\textbf{g})~aortic arch and (\textbf{h})~three vessel views.
(\textbf{d})~A line profile at the intersection of the four chamber and mid-short axis views (dashed yellow lines) shows the contraction and dilation of the ventricles with cardiac phase~($\vartheta$).
Ventriculo-arterial connections can be seen in outflow tract views with the pulmonary artery~(PA) from the right ventricle~(RV) in (e) and the aorta~(Ao) arising from the left ventricle~(LV) in (f). 
Systemic venous connections of the superior~(SVC) and inferior~(IVC) vena cava with the right atrium~(RA) can also be seen in (f). 
The ductal arch~(DA) can be seen in both (e) and (h), connecting the PA to the descending aorta~(DAo), while the Ao arch can be seen in (g) and (h). 
All boxes bounding the re-sliced views measure 65×65 mm. 
The fetal heart is shown in radiological orientation, i.e., image axes up and right relative to the page correspond to left, anterior and/or superior anatomical directions. 
Views are shown using spatial B-spline interpolation to avoid voxel distortion.}
\label{fig:id09_cine_vol_views}
\end{figure}

Reconstructed 4D cine volumes could be viewed in any arbitrary plane and at any phase in the cardiac cycle to examine cardiovascular morphology and connectivity.
An example is shown in Figure~\ref{fig:id09_cine_vol_views} where the volume is viewed in variety of double-oblique cardiac planes.
The relative size of the chambers of the heart can be seen in horizontal long axis~(four chamber) and mid-short axis views, with dilation and contraction of the left~(LV) and right~(RV) ventricles and left~(LA) and right~(RA) atria as the heart beats.
Outflow tract and other off-axis views reveal arterial and venous connections of the aorta~(Ao), pulmonary artery~(PA) and the  superior~(SVC) and inferior~(IVC) vena cava, as well as the positions of the Ao and ductal arch~(DA). 
A video of the reconstructed 4D volume (Video S2) and intermediate results for motion-correction, cardiac synchronisation and outlier rejection~(Figure~\ref{fig:id09_rr_v_time}) for this case are included in the Supporting Information.

\begin{figure}[htb]
\centering
\includegraphics[width = 0.95\textwidth]{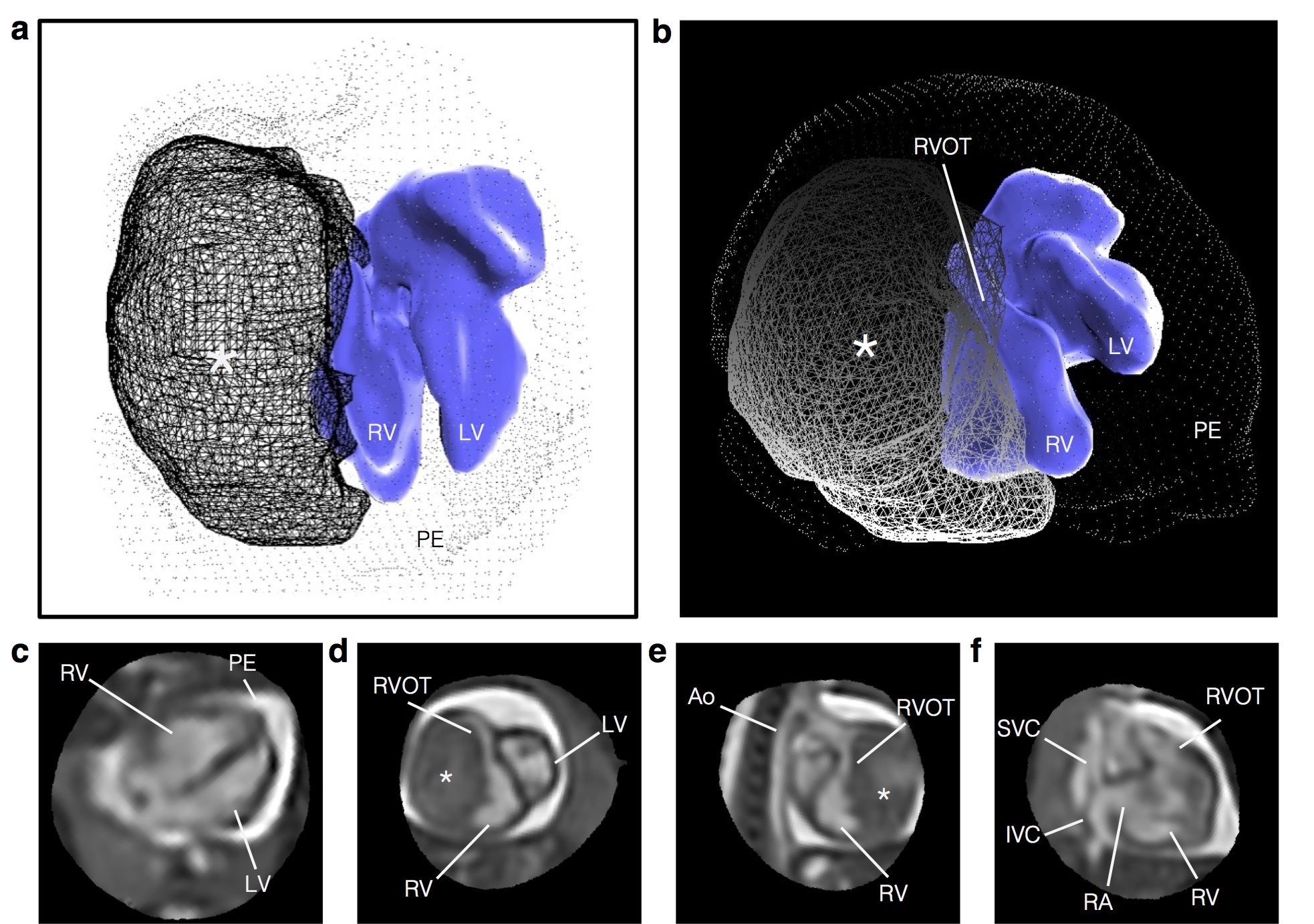}
\caption[4D Cine MRI Volume in Fetus with Cardiac Fibroma]{Reconstructed 4D cine volume of the heart of a 33\textsuperscript{+3} week gestational age fetus~(ID11) with large fibroma, measuring 24×24×37~mm, involving the lateral wall of the right ventricle. 
A segmented volume rendering is shown in (\textbf{a}) superior and (\textbf{b}) anterior views, for reference of relative shape and position of the cardiac blood pool~(blue surface), fibroma~(lined mesh) and pericardial effusion~(dotted mesh).
The reconstructed volume is shown in (\textbf{c})~four chamber, (\textbf{d})~short axis, (\textbf{e})~right ventricular outflow tract and (\textbf{f})~right three chamber views. 
Extensive associated pericardial effusion~(PE) can be seen as bright signal surrounding the heart.  
The fibroma~(asterisk) is encapsulated within myocardial tissue and confined to the margin of the right ventricular~(RV) cavity, causing external compression of the right side of the heart, including the RV, right atrium~(RA) and right ventricular outflow tract~(RVOT), while the left ventricle~(LV), aorta~(Ao), and both the superior~(SVC) and inferior~(IVC) vena cava appear normal. 
All boxes bounding the re-sliced views in (c--f) measure 80×80~mm and the fetal heart is shown in radiological orientation, i.e., image axes towards the top and right of the page correspond to left, anterior and superior anatomical directions. 
Views are shown using spatial B-spline interpolation to avoid voxel distortions.}
\label{fig:id11_cine_vol_views}
\end{figure}

Figure~\ref{fig:id11_cine_vol_views} shows a fetus with a cardiac tumour~(ID11). 
The ability to visualise the relationship of the mass and the myocardium in both space and time allowed for qualitative assessment of the impact of the tumour on cardiac function.
A particularly large outlier class was estimated in this case~(44\% of frames rejected) compared to all other cases in cohort~1~(median 10\% rejected, range 5--28\%), however this particular heart was grossly different in shape and appearance compared to the other fetal cases.
It may be that the current implementation of outlier rejection is better suited to fine, high-contrast anatomy, such as the normal fetal brain and heart, than to anatomy containing large regions with homogeneous signal.

\begin{figure}[htb]
\centering
\bigskip
\includegraphics[width = 1.0\textwidth]{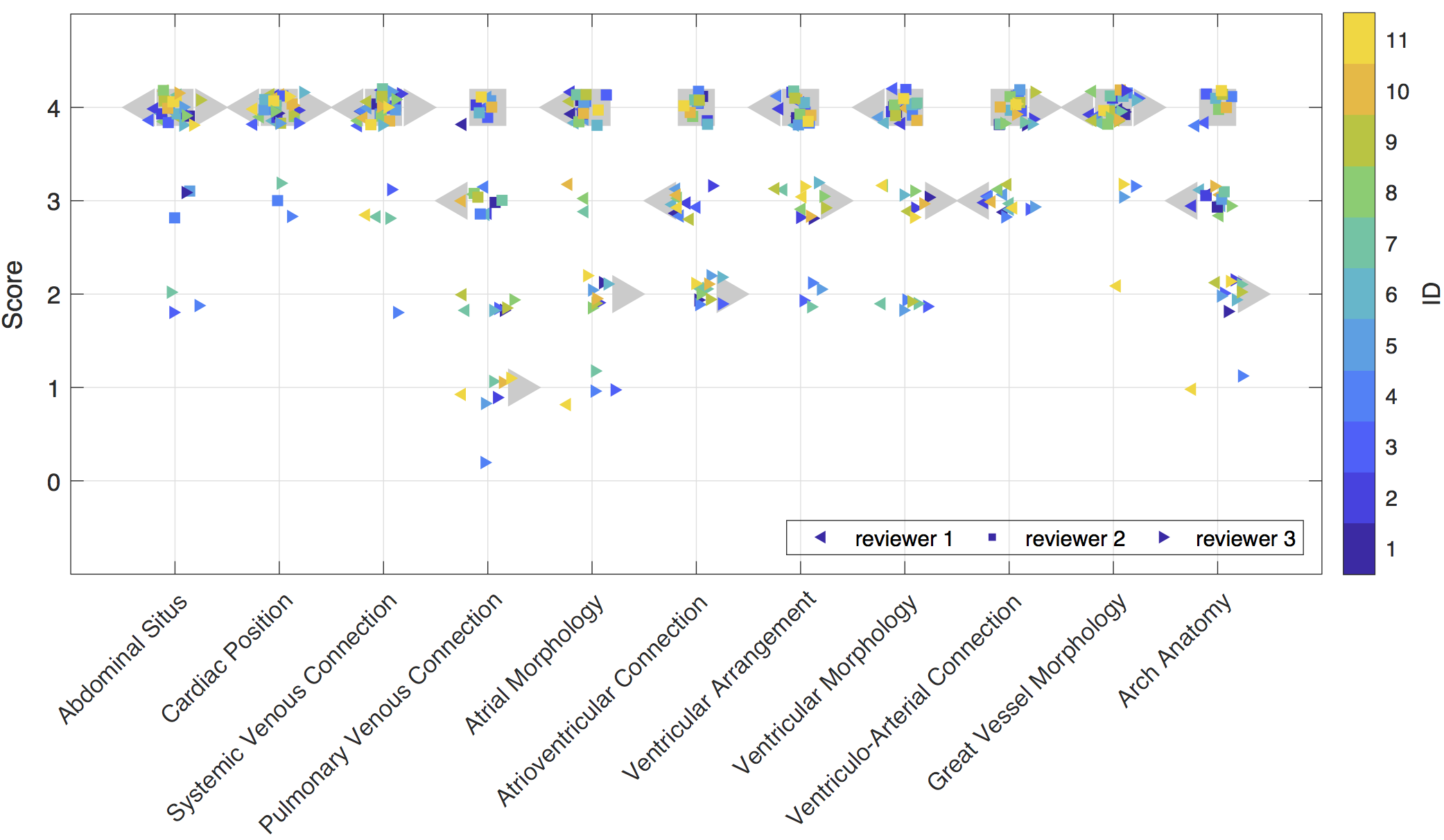}
\caption[Reviewer Score by Evaluation Category]{Reviewer score by evaluation category for cardiac cine volumes of eleven fetuses in cohort~1. Scores are shown for reviewer 1 (left-pointing triangles), reviewer 2 (squares), and reviewer 3 (right-pointing triangles), with four, seven, and one years experience reading cardiac MRI, respectively. Median reviewer scores are shown as large grey markers while coloured markers denote scores for individual fetal cases.}
\label{fig:reviewer_score_v_category}
\end{figure}

The expert evaluation is summarised in Figure~\ref{fig:reviewer_score_v_category} for fetal cases in cohort~1.
Scores were generally high in all categories suggesting a potential to assess a range of cardiac anatomy in the cine volumes.
Mean scores were between~3 and 4 for all fetal cases~(median~3.4) in cohort~1~(Table~\ref{tab:fetal_results}) indicating that these cine volumes in were of adequate quality to determine most anatomical details.
The lowest mean scores across all reviewers were for pulmonary venous connections, atrioventricular connections, and arch anatomy, suggesting that small, complex anatomical features were the most challenging for the reviewers.
The reconstructed volumes of the second cohort of fetal subjects were visually assessed by one reviewer~(DL) and determined to be of similar high quality as the first cohort in the majority of cases, while insufficient data resulted in reduced quality in two cases for which 1920~(ID16) and 2590~(ID18) image frames were acquired covering the fetal heart, compared to a  minimum of 3072 frames for all cases in cohort~1.

All steps of the proposed method were performed as described, with the exception of two cases in cohort~1 that required some motion-corrupted data to be manually excluded prior to reconstruction~(ID04,ID07).
Prior to excluding motion-corrupted slices these two cases had ${\text{dev}(\mathbf{A})}=$~3.8~mm~(ID04) and 2.5~mm~(ID07), larger than than all other cases in cohort~1~(median 1.3~mm, range 1.0--2.3~mm).
Manual intervention resulted in ${\text{dev}(\mathbf{A})}=$~1.8~mm~(ID04) and 2.3~mm~(ID07), and yielded visually improved 4D cine reconstructions in both cases, though both received the low mean reviewer scores and had large outlier classes, as expected. 
Results from motion-corrupted data are shown in Supporting Figure~\ref{fig:id04_realtime_and_cine_images}.

\begin{SCfigure}[][htb]
\centering
\includegraphics[width = 0.62\textwidth]{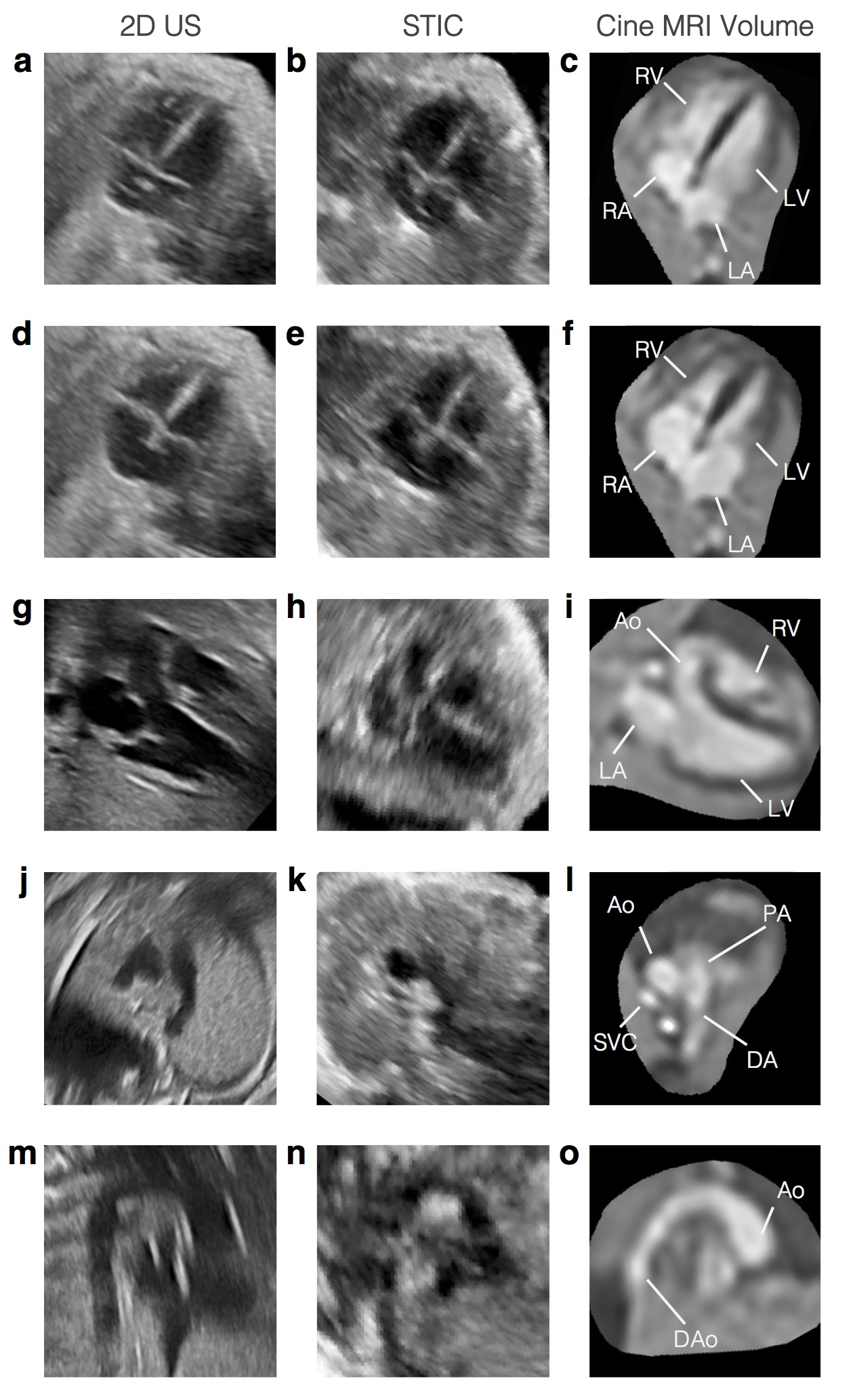}
\caption[Comparison of Ultrasound and MRI Depiction of Fetal Cardiovascular Anatomy]{Comparison of MRI and ultrasound~(US) in a 32\textsuperscript{+1} week gestational age fetus~(ID01) with moderate dilation of the ascending aortic root and rightward tortuosity of the proximal ascending aorta. The fetal heart is shown in matched views in 2D echocardiogram~(left column), 4D~spatio-temporal image correlation~(STIC) ultrasound (centre column) and reconstructed 4D cine MRI volumes~(right column). Four chamber views at \mbox{(\textbf{a})-(\textbf{c})}~end-ventricular diastole and \mbox{(\textbf{d})-(\textbf{f})}~end-ventricular systole show the alternating contraction and expansion of the ventricles~(LV,RV) and atria~(LA,RA), with balanced chambers. \mbox{(\textbf{g})-(\textbf{i})}~Left three chamber view, \mbox{(\textbf{j})-(\textbf{l})}~three vessel view and \mbox{(\textbf{m})-(\textbf{o})}~aortic arch in sagittal view revealing a dilated aorta~(Ao) at the sinotubular junction, compared to the pulmonary artery~(PA), ductal arch~(DA) and descending aorta~(DAo). 
Some artefacts can be seen in the 2D~US images and STIC volumes, occasionally obscuring cardiac anatomy, while MRI had good signal coverage of the entire cardiovascular system.
All bounding boxes measuring 65×65 mm and the fetal heart is shown in radiological orientation. The 4D MRI views are shown using spatial B-spline interpolation to avoid voxel distortions.}
\label{fig:id01_cine_v_us_v_stic}
\end{SCfigure}

Cine volumes were compared with 2D and STIC ultrasound in the cases where matched data was acquired.
An example is shown in Figure~\ref{fig:id01_cine_v_us_v_stic} where the three imaging methods are compared in matched views. 
Though STIC quality has been reported to be reduced in fetuses in the third trimester~\cite{Zhao2016}, the STIC volume of this 32\textsuperscript{+1}~week gestational age fetus was the best quality of those acquired. 
In the other six fetal cases with matched ultrasound data, 2D echocardiography images were of comparable quality to those shown.
However STIC volumes in four cases were clearly of poor quality, particularly in the through-plane direction, presumably due to fetal motion, while MRI 4D reconstructions were of high quality in all seven cases with matched US data.

Both 2D~US and STIC showed clear definition of vessel and chamber boundaries, including valves, with high spatial and temporal resolution. 
Reconstructed cine MRI volumes had good contrast between blood and surrounding tissue.
However, visualisation of valves and other small and rapidly moving anatomy was limited due to the spatio-temporal resolution of the acquisition.
While the real-time aspect of 2D~US made it robust to fetal motion, interpretation of complex anatomy was limited to the views acquired at the time of examination. 
By contrast, both STIC and MRI cine volumes allowed for offline analysis in arbitrary planes with full coverage of the heart and great vessels, facilitating understanding of the spatial relationships between cardiovascular structures. 

\begin{figure}[htb]
\centering
\includegraphics[width = 0.75\textwidth]{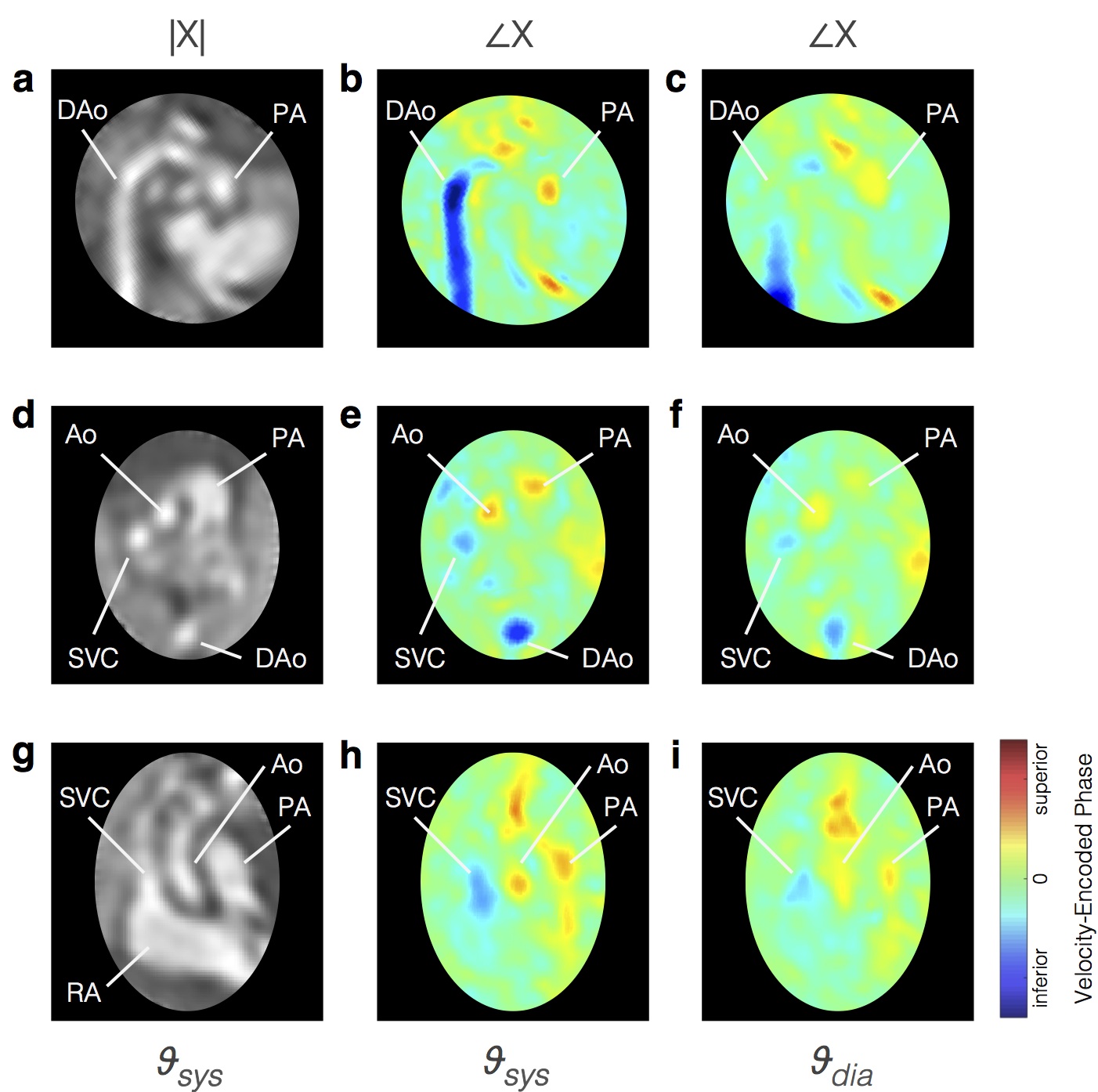}
\caption[Psuedo 4D Velocity Mapping in a Healthy Fetus]{Pseudo 4D velocity mapping in a healthy 30\textsuperscript{+3} week gestational age fetus~(ID02). Complex-valued cine volume magnitude, $|\mathbf{X}|$, and velocity-encoded phase, $\angle \mathbf{X}$, are shown in (\textbf{a})-(\textbf{c})~sagittal aortic arch plane, (\textbf{d})-(\textbf{f})~three vessel view and (\textbf{g})-(\textbf{i})~coronal plane perpendicular to three vessel view. The phase  of the reconstructed cine volume was most sensitive to velocities in the fetal superior-inferior direction, perpendicular to the target stack acquired transverse to the fetal trunk. Flow in the inferior direction~(blue), is seen in the descending aorta~(DAo) and superior vena cava~(SVC), while flow in the superior direction~(red) is seen in the ascending aorta~(Ao) and pulmonary artery~(PA). Peak flows can be seen in systole,~$\vartheta_{sys}$, with reduced flow during diastole,~$\vartheta_{dia}$. The right atrium~(RA) is labelled in (g) for reference. The fetal heart is oriented in radiological convention with double-oblique planes shown using spatial B-spline interpolation to avoid voxel distortion.}
\label{fig:id02_velocity_map}
\end{figure}

A complex-valued 4D cine volume is shown in Figure~\ref{fig:id02_velocity_map} with the velocity-encoded phase most sensitive to flow in the fetal superior-inferior direction and the magnitude-valued volume shown as an anatomical reference. 
High flow can be seen in the great vessels in systole, with reduced flow in diastole.

\section{Discussion}  

Whole-heart 4D reconstruction of the fetal heart from multi-planar dynamic MRI was successful using the proposed framework in all fetal cases with at least 3000 image frames covering the heart.
These reconstructed volumes could be visualised in any 2D plane without the need for highly-specific scan plane prescription prior to acquisition or maternal breath hold to minimise motion.
The three-dimensional spatial aspect of these volumes allowed for interpretation of the complex connections of the cardiac chambers and vessels, while the temporal aspect improved the depiction of pulsatile anatomy. 

Reconstructions of simulated MR images confirmed that spatial and temporal features could be reliably recovered, within the limits of the resolution of the acquisition.
There was some blurring in the cine volumes reconstructed from simulated MR images compared to the high resolution ground truth, which can be attributed to the low spatio-temporal resolution sampling of the dynamic data. 
There was also a ringing-like signal intensity effect in the cine volumes that was apparent in the simulation results~(Fig.~\ref{fig:sim_mri}d), but had a more subtle impact on the fetal results.
This effect may arise from the approximated in-plane spatial PSF, however testing is not straightforward due to the complexity of  implementing and computing a sinc PSF.

Overall high scores were assigned by three expert readers in an evaluation based on the segmental approach to defining cardiac anatomy and pathology.
The lowest scores were assigned in categories focused on small, complex anatomical features, such as pulmonary venous connections and arch anatomy, where spatio-temporal resolution limited the depiction of fine details, as was also the case for small septal defects and all atrioventricular and ventriculo-arterial valves. 
Though a category such as ventricular morphology received high scores in this evaluation, as there was enough other information to give the reviewers high confidence, particular abnormalities, such as valvular defects, were more visible in ultrasound images.
However, some cardiac anatomy was obscured by artefact in the ultrasound images due to acoustic shadowing and, in the case of STIC volumes, blurring and misalignment.
The results of the expert evaluation suggest that there is potential utility of the method to generate 4D volumes that can be used for a comprehensive assessment of the fetal heart, either as an adjunct to ultrasound or in combination with other MRI techniques. 
For example, while aortic and ductal arch branching patterns may be difficult to determine in the 4D reconstruction, a static volume reconstruction of T\textsubscript{2}-weighted spin echo MRI~\cite{Lloyd2018} may more clearly depict the details of the arch anatomy.
However, 4D reconstruction captures cardiac movement across the entire cardiac cycle, which may improve visualisation of anatomy affected by cardiac motion.

The proposed framework is fully automated aside from anatomical ROIs and identification of a target for initial stack-stack registration.
While these user-interactions are not particularly laborious, they could be replaced with automated techniques for segmentation~\cite{KeraudrenMICCAI2015,Demesmaeker2017} and target stack identification using a measure of slice alignment~\cite{Kainz2015}.
Manual intervention was required to exclude data with significant motion in two cases in cohort~1. 
In the original cine volumes reconstructed using these motion-corrupted data the anatomy was blurred and inconsistent as errors in stack- and slice-wise motion-correction lead to slice-slice cardiac synchronisation errors, and the reconstruction did not converge to a good depiction of the underlying fetal heart.
A motion metric~\cite{Kainz2015} could be used in a preprocessing stage to identify slices with significant motion prior to any motion correction or cardiac synchronisation without need for intervention.
The focus of this work was on volumetric depiction of the whole fetal heart and, consequently, the acquisition and reconstruction of k-t SENSE MRI was not fundamentally changed from the 2D implementation~\cite{vanAmerom2017} and many of the potential improvements previously suggested still apply. 

The use of a constant heart rate for each slice does not accommodate beat-to-beat variation resulting in small timing errors~\cite{vanAmerom2017}. 
Cardiac monitoring using doppler ultrasound gating~\cite{Kording2018} or image-based self-gating~\cite{Demesmaeker2017,Larson2004,Nijm2008} could reduce these timing errors, either in combination with or instead of the current approach.

Simulation experiments showed slice-slice cardiac synchronisation error of less than 5\% of the cardiac cycle for the range of displacements measured in the majority of fetal cases (Fig.~\ref{fig:simulation_results}b), suggesting the proposed method provided reliable results.
Cardiac synchronisation was performed prior cine volume reconstruction in the proposed framework and synchronisation errors may be improved if the two steps were interleaved in a similar manner to frame-volume registration. 

Outlier rejection was performed as implemented by Kuklisova \textit{et al.}~\cite{Kuklisova-Murgasova2012}, with robust statistics shown to effectively reduce the influence of inconsistent data. 
As magnitude-valued images were used, sensitivity to image artefacts that manifest in the phase of complex-valued images was reduced compared to the complex-valued outlier rejection used in the 2D framework~\cite{vanAmerom2017}.
However, this did not appear to have a dramatic effect as dynamic images were reconstructed using spatially-uniform k-t SENSE regularisation, thereby suppressing image artefact to some extent.
Complex-valued outlier rejection should be possible in a velocity-sensitive reconstruction framework.

\section{Conclusion}  

In summary, four-dimensional representation of the fetal heart and great vessels was achieved using a highly-accelerated multi-planar dynamic MRI acquisition combined with retrospective motion correction, cardiac synchronisation, outlier rejection and volumetric cine reconstruction in the image domain.
The motion-tolerant framework did not require maternal breath-hold or precise scan planning during acquisition, and reconstruction was fully automated aside from user-specified fetal heart and chest ROIs.
The framework proved to be robust when used on fetal data and successfully generated good quality cine volumes in all fetal cases where sufficient data was collected covering the entire fetal heart. 
Experts had an overall high confidence in a comprehensive evaluation of the fetal heart using the reconstructed cine volumes.

Reconstructions from simulated MR images confirmed that correct spatial and temporal features could be reliably recovered, though there was some blurring due to the spatio-temporal resolution of the acquired dynamic images. 
The use of image-domain motion correction methods conferred significant advantages in terms of outlier rejection and the ability to accommodate quite large fetal motion, but had the downside of presenting major challenges for increasing the spatial and temporal resolution of the dynamic MR acquisition.

The proposed methods show promise as a framework for motion-corrected reconstruction and 4D assessment of the fetal heart and great vessels. 
Promising results from a preliminary assessment of velocity-sensitive volume reconstruction suggest potential for simultaneous reconstruction of a four-dimensional cine and fully-encoded velocity volumes from multi-planar dynamic bSSFP MRI.

\FloatBarrier

\section*{Acknowledgements}

Thank you to Joanna Allsop, Ana Gomes and Elaine Green for their oversight during the scanning of volunteers and patients, and to all the obstetricians and administrators that made this work possible.
We also gratefully acknowledge Torben Schneider and Jouke Smink for their input with pulse sequence programming, Bernhard Kainz and Tong Zhang for their insight in to MRI acquisition modelling and Camila Munoz for her assistance with MRXCAT.
This work received funding from the Engineering and Physical Sciences Research Council~(EPSRC) [EP/H046410/1] and the Medical Research Council Strategic Fund [MR/K0006355/1], and was supported by the Wellcome Trust IEH Award intelligent Fetal Imaging and Diagnosis~(iFIND) project [102431], the Wellcome/EPSRC Centre for Medical Engineering [WT~203148/Z/16/Z] and the National Institute of Health Research~(NIHR) Biomedical Research Centre based at Guy’s and St Thomas’ National Health Services~(NHS) Foundation Trust and King’s College London.
The views expressed are those of the authors and not necessarily those of the NHS, the NIHR or the Department of Health.

\section*{Supporting Information}

\noindent\textbf{Figure~\ref{fig:simulation_nrmse_v_delta_v_lambda}.} Optimisation and evaluation using simulated MR images.
(\textbf{a})~Reconstruction parameter optimisation.
(\textbf{b--c})~Optimisation and (\textbf{d--e})assessment of 4D reconstruction.

\noindent\textbf{Video~S2.} Animation demonstrating spatial and temporal aspects of 4D whole-heart reconstruction of healthy 28\textsuperscript{+0}~week gestational age fetus~(ID09) shown in Figure~\ref{fig:id09_cine_vol_views}, with volume rendering of blood pool (red) for reference.

\noindent\textbf{Figure~\ref{fig:id09_rr_v_time}.} Example cardiac synchronisation, motion-correction and outlier rejection results in healthy 28\textsuperscript{+0}~week gestational age fetus~(ID09) shown in Figure~\ref{fig:id09_cine_vol_views}.  
(\textbf{a})~Estimated heart rates. 
(\textbf{b})~Estimated spatial transformations and image frame-wise probabilities.
(\textbf{c--d})~Outlier rejection in the presence of motion.
(\textbf{e})~Voxel- and (\textbf{f}) frame-wise robust statistics. 

\noindent\textbf{Figure~\ref{fig:id04_realtime_and_cine_images}.} Significant movement of 29\textsuperscript{+6}~week gestational age fetus~(ID04).
(\textbf{a--b})~Dynamic MR images and (\textbf{c--d})~impact of fetal movement on reconstructed cine volume.

\printendnotes

\bibliography{mendeley}

\begin{thebibliography}{35}
\providecommand{\natexlab}[1]{#1}
\providecommand{\url}[1]{\texttt{#1}}
\providecommand{\urlprefix}{}

\bibitem[{Votino et~al.(2012)Votino, C and Jani, J and Damry, N and Dessy, H
  and Kang, X and Cos, T and Divano, L and Foulon, W and De Mey, J and Cannie,
  M}]{Votino2012}
Votino C, Jani J, Damry N, Dessy H, Kang X, Cos T, et~al.
\newblock {Magnetic resonance imaging in the normal fetal heart and in
  congenital heart disease.}
\newblock Ultrasound in Obstetrics {\&} Gynecology 2012 3;39(3):322--9.
\newblock
  \urlprefix\url{https://obgyn.onlinelibrary.wiley.com/doi/abs/10.1002/uog.10061}.

\bibitem[{Donofrio et~al.(2014)Donofrio, Mary T and Moon-Grady, Anita J and
  Hornberger, Lisa K and Copel, Joshua a and Sklansky, Mark S and Abuhamad,
  Alfred and Cuneo, Bettina F and Huhta, James C and Jonas, Richard a and
  Krishnan, Anita and Lacey, Stephanie and Lee, Wesley and Michelfelder, Erik C
  and Rempel, Gwen R and Silverman, Norman H and Spray, Thomas L and
  Strasburger, Janette F and Tworetzky, Wayne and Rychik, Jack}]{Donofrio2014}
Donofrio MT, Moon-Grady AJ, Hornberger LK, Copel Ja, Sklansky MS, Abuhamad A,
  et~al.
\newblock {Diagnosis and Treatment of Fetal Cardiac Disease: A Scientific
  Statement From the American Heart Association.}
\newblock Circulation 2014
  4;\urlprefix\url{https://www.ahajournals.org/doi/abs/10.1161/01.cir.0000437597.44550.5d}.

\bibitem[{Roy et~al.(2017)Roy, Christopher W and Seed, Mike and Kingdom, John C
  and Macgowan, Christopher K}]{Roy2017_MocoCineFetalCMR}
Roy CW, Seed M, Kingdom JC, Macgowan CK.
\newblock {Motion compensated cine CMR of the fetal heart using radial
  undersampling and compressed sensing}.
\newblock Journal of Cardiovascular Magnetic Resonance 2017;19(1):29.
\newblock
  \urlprefix\url{https://jcmr-online.biomedcentral.com/articles/10.1186/s12968-017-0346-6}.

\bibitem[{van Amerom et~al.(2017)van Amerom, Joshua F.P. and Lloyd, David F.A.
  and Price, Anthony N. and Kuklisova Murgasova, Maria and Aljabar, Paul and
  Malik, Shaihan J. and Lohezic, Maelene and Rutherford, Mary A. and
  Pushparajah, Kuberan and Razavi, Reza and Hajnal, Joseph V.}]{vanAmerom2017}
van Amerom JFP, Lloyd DFA, Price AN, Kuklisova~Murgasova M, Aljabar P, Malik
  SJ, et~al.
\newblock {Fetal cardiac cine imaging using highly accelerated dynamic MRI with
  retrospective motion correction and outlier rejection}.
\newblock Magnetic Resonance in Medicine 2017 1;79(1):327--338.
\newblock \urlprefix\url{http://doi.wiley.com/10.1002/mrm.26686}.

\bibitem[{Kuklisova-Murgasova et~al.(2012)Kuklisova-Murgasova, Maria and
  Quaghebeur, Gerardine and Rutherford, Mary A. and Hajnal, Joseph V. and
  Schnabel, Julia A.}]{Kuklisova-Murgasova2012}
Kuklisova-Murgasova M, Quaghebeur G, Rutherford MA, Hajnal JV, Schnabel JA.
\newblock {Reconstruction of fetal brain MRI with intensity matching and
  complete outlier removal.}
\newblock Medical Image Analysis 2012 12;16(8):1550--64.
\newblock
  \urlprefix\url{http://www.sciencedirect.com/science/article/pii/S1361841512000965}.

\bibitem[{Gholipour et~al.(2010)Gholipour, Ali and Estroff, Judy a and
  Warfield, Simon K}]{Gholipour2010}
Gholipour A, Estroff Ja, Warfield SK.
\newblock {Robust super-resolution volume reconstruction from slice
  acquisitions: application to fetal brain MRI.}
\newblock IEEE Transactions on Medical Imaging 2010 10;29(10):1739--58.
\newblock \urlprefix\url{https://dx.doi.org/10.1109%2FTMI.2010.2051680}.

\bibitem[{Roy et~al.(2017)Roy, Christopher W. and Seed, Mike and Macgowan,
  Christopher K.}]{Roy2017_CSMOG}
Roy CW, Seed M, Macgowan CK.
\newblock {Accelerated MRI of the fetal heart using compressed sensing and
  metric optimized gating}.
\newblock Magnetic Resonance in Medicine 2017 6;77(6):2125--2135.
\newblock \urlprefix\url{http://doi.wiley.com/10.1002/mrm.26290}.

\bibitem[{Chaptinel et~al.(2017)Chaptinel, Jerome and Yerly, Jerome and
  Mivelaz, Yvan and Prsa, Milan and Alamo, Leonor and Vial, Yvan and Berchier,
  Gregoire and Rohner, Chantal and Gudinchet, François and Stuber,
  Matthias}]{Chaptinel2017}
Chaptinel J, Yerly J, Mivelaz Y, Prsa M, Alamo L, Vial Y, et~al.
\newblock {Fetal cardiac cine magnetic resonance imaging in utero}.
\newblock Scientific Reports 2017 12;7(1):15540.
\newblock \urlprefix\url{http://www.nature.com/articles/s41598-017-15701-1}.

\bibitem[{Haris et~al.(2016)Haris, Kostas and Hedstrom, Erik and Bidhult,
  Sebastian and Testud, Frederik and Kantasis, George and Engblom, Henrik and
  Carlsson, Marcus and Maglaveras, Nicos and Heiberg, Einar and Hansson, Stefan
  R and Arheden, Hakan and Aletras, Anthony H}]{Haris2016FetalIGRASP}
Haris K, Hedstrom E, Bidhult S, Testud F, Kantasis G, Engblom H, et~al.
\newblock {Fetal Cardiac MRI with self-gated iGRASP}.
\newblock In: International Society for Magnetic Resonance in Medicine; 2016.
  p. 3104.

\bibitem[{van Amerom et~al.(2018)van Amerom, Joshua F P and Lloyd, David F A
  and Murgasova, Maria Kuklisova and Price, Anthony N and Malik, Shaihan J and
  Poppel, Milou Van and Pushparajah, Kuberan and Rutherford, Mary A and Razavi,
  Reza and Hajnal, Joseph V}]{vanAmeromISMRM2018}
van Amerom JFP, Lloyd DFA, Murgasova MK, Price AN, Malik SJ, Poppel MV, et~al.
\newblock {Fetal whole-heart 3D cine reconstruction using motion-corrected
  multi-slice dynamic imaging}.
\newblock In: International Society for Magnetic Resonance in Medicine; 2018.
  p. 1052.

\bibitem[{Tsao et~al.(2003)Tsao, Jeffrey and Boesiger, Peter and Pruessmann,
  Klaas P}]{Tsao2003}
Tsao J, Boesiger P, Pruessmann KP.
\newblock {k-t BLAST and k-t SENSE: dynamic MRI with high frame rate exploiting
  spatiotemporal correlations}.
\newblock Magnetic Resonance in Medicine 2003 11;50(5):1031--42.
\newblock \urlprefix\url{https://doi.org/10.1002/mrm.10611}.

\bibitem[{Jiang et~al.(2007)Jiang, Shuzhou and Member, Student and Xue, Hui and
  Glover, Alan and Rutherford, Mary and Rueckert, Daniel and Hajnal, Joseph
  V.}]{Jiang2007}
Jiang S, Member S, Xue H, Glover A, Rutherford M, Rueckert D, et~al.
\newblock {MRI of moving subjects using multislice Snapshot images with Volume
  Reconstruction (SVR): Application to fetal, neonatal, and adult brain
  studies}.
\newblock IEEE Transactions on Medical Imaging 2007;26(7):967--980.
\newblock \urlprefix\url{https://doi.org/10.1109/TMI.2007.895456}.

\bibitem[{Rousseau et~al.(2006)Rousseau, Francois and Glenn, Orit A and
  Iordanova, Bistra and Rodriguez-Carranza, Claudia and Vigneron, Daniel B and
  Barkovich, James A and Studholme, Colin}]{Rousseau2006}
Rousseau F, Glenn OA, Iordanova B, Rodriguez-Carranza C, Vigneron DB, Barkovich
  JA, et~al.
\newblock {Registration-based approach for reconstruction of high-resolution in
  utero fetal MR brain images.}
\newblock Academic Radiology 2006 9;13(9):1072--81.
\newblock
  \urlprefix\url{http://www.sciencedirect.com/science/article/pii/S1076633206002753}.

\bibitem[{Aljabar et~al.(2008)Aljabar, P and Bhatia, K K and Murgasova, M and
  Hajnal, J V and Boardman, J P and Srinivasan, L and Rutherford, M A and Dyet,
  L E and Edwards, A D and Rueckert, D}]{Aljabar2008}
Aljabar P, Bhatia KK, Murgasova M, Hajnal JV, Boardman JP, Srinivasan L, et~al.
\newblock {Assessment of brain growth in early childhood using
  deformation-based morphometry.}
\newblock NeuroImage 2008 1;39(1):348--58.
\newblock
  \urlprefix\url{http://www.sciencedirect.com/science/article/pii/S1053811907006878}.

\bibitem[{Rousseeuw and Croux(1993)Rousseeuw, Peter J. and Croux,
  Christophe}]{Rousseeuw1993}
Rousseeuw PJ, Croux C.
\newblock {Alternatives to the Median Absolute Deviation}.
\newblock Journal of the American Statistical Association 1993
  12;88(424):1273--1283.
\newblock
  \urlprefix\url{http://www.tandfonline.com/doi/abs/10.1080/01621459.1993.10476408}.

\bibitem[{Hand et~al.(2010)Hand, J W and Li, Y and Hajnal, J V}]{Hand2010}
Hand JW, Li Y, Hajnal JV.
\newblock {Numerical study of RF exposure and the resulting temperature rise in
  the foetus during a magnetic resonance procedure.}
\newblock Physics in Medicine and Biology 2010 2;55(4):913--30.
\newblock \urlprefix\url{https://doi.org/10.1088/0031-9155/55/4/001}.

\bibitem[{Glover et~al.(1995)Glover, P and Hykin, J and Gowland, P and Wright,
  J and Johnson, I and Mansfield, P}]{Glover1995}
Glover P, Hykin J, Gowland P, Wright J, Johnson I, Mansfield P.
\newblock {An assessment of the intrauterine sound intensity level during
  obstetric echo-planar magnetic resonance imaging.}
\newblock The British journal of radiology 1995 10;68(814):1090--4.
\newblock
  \urlprefix\url{http://www.birpublications.org/doi/10.1259/0007-1285-68-814-1090}.

\bibitem[{Tsao et~al.(2005)Tsao, Jeffrey and Kozerke, Sebastian and Boesiger,
  Peter and Pruessmann, Klaas P}]{Tsao2005}
Tsao J, Kozerke S, Boesiger P, Pruessmann KP.
\newblock {Optimizing spatiotemporal sampling for k-t BLAST and k-t SENSE:
  application to high-resolution real-time cardiac steady-state free
  precession.}
\newblock Magnetic resonance in medicine 2005 6;53(6):1372--82.
\newblock \urlprefix\url{https://doi.org/10.1002/mrm.20483}.

\bibitem[{Deimling and Heid(1994)Deimling, M. and Heid, O.}]{Deimling1994}
Deimling M, Heid O.
\newblock {Magnetization Prepared True FISP Imaging}.
\newblock In: International Society for Magnetic Resonance in Medicine,
  vol.~22; 1994. p. 495.

\bibitem[{DeVore et~al.(2003)DeVore, G. R. and Falkensammer, P. and Sklansky,
  M. S. and Platt, L. D.}]{DeVore2003}
DeVore GR, Falkensammer P, Sklansky MS, Platt LD.
\newblock {Spatio-temporal image correlation (STIC): new technology for
  evaluation of the fetal heart.}
\newblock Ultrasound in Obstetrics and Gynecology 2003 10;22(4):380--7.
\newblock \urlprefix\url{http://doi.wiley.com/10.1002/uog.217}.

\bibitem[{Yoo et~al.(1999)Yoo, S J and Lee, Y H and Cho, K S and Kim, D
  Y}]{Yoo1999}
Yoo SJ, Lee YH, Cho KS, Kim DY.
\newblock {Sequential segmental approach to fetal congenital heart disease.}
\newblock Cardiology in the young 1999 7;9(4):430--44.
\newblock \urlprefix\url{https://doi.org/10.1017/S1047951100005266}.

\bibitem[{Carvalho et~al.(2005)Carvalho, J. S. and Ho, S. Y. and Shinebourne,
  E. A.}]{Carvalho2005}
Carvalho JS, Ho SY, Shinebourne EA.
\newblock {Sequential segmental analysis in complex fetal cardiac
  abnormalities: a logical approach to diagnosis}.
\newblock Ultrasound in Obstetrics and Gynecology 2005 8;26(2):105--111.
\newblock \urlprefix\url{http://doi.wiley.com/10.1002/uog.1970}.

\bibitem[{Wissmann et~al.(2014)Wissmann, Lukas and Santelli, Claudio and
  Segars, William Paul and Kozerke, Sebastian}]{Wissmann2014}
Wissmann L, Santelli C, Segars WP, Kozerke S.
\newblock {MRXCAT: Realistic Numerical Phantoms for Cardiac Magnetic Resonance
  Imaging}.
\newblock Journal of Cardiovascular Magnetic Resonance 2014 8;16(1):63.
\newblock \urlprefix\url{http://jcmr-online.com/content/16/1/63}.

\bibitem[{Alexa et~al.(2002)Alexa, Marc and {Marc} and {Alexa} and
  {Marc}}]{Alexa2002}
Alexa M, {Marc}, {Alexa}, {Marc}.
\newblock {Linear combination of transformations}.
\newblock In: Computer Graphics and Interactive Techniques, vol.~21 ACM Press;
  2002. p. 380.
\newblock \urlprefix\url{https://doi.org/10.1145/566570.566592}.

\bibitem[{Markl et~al.(2003)Markl, M. and Alley, M.T. and Pelc,
  N.J.}]{Markl2003}
Markl M, Alley MT, Pelc NJ.
\newblock {Balanced phase-contrast steady-state free precession (PC-SSFP): A
  novel technique for velocity encoding by gradient inversion}.
\newblock Magnetic Resonance in Medicine 2003 5;49(5):945--952.
\newblock \urlprefix\url{http://doi.wiley.com/10.1002/mrm.10451}.

\bibitem[{Markl and Pelc(2004)Markl, Michael and Pelc, Norbert J.}]{Markl2004}
Markl M, Pelc NJ.
\newblock {On flow effects in balanced steady-state free precession imaging:
  Pictorial description, parameter dependence, and clinical implications}.
\newblock Journal of Magnetic Resonance Imaging 2004 10;20(4):697--705.
\newblock \urlprefix\url{http://doi.wiley.com/10.1002/jmri.20163}.

\bibitem[{Nielsen and Nayak(2009)Nielsen, Jon-Fredrik and Nayak, Krishna
  S.}]{Nielsen2009}
Nielsen JF, Nayak KS.
\newblock {Referenceless phase velocity mapping using balanced SSFP}.
\newblock Magnetic Resonance in Medicine 2009 5;61(5):1096--1102.
\newblock \urlprefix\url{http://doi.wiley.com/10.1002/mrm.21884}.

\bibitem[{Zhao et~al.(2016)Zhao, Liqing and Wu, Yurong and Chen, Sun and Ren,
  Yunyun and Chen, Ping and Niu, Jianmei and Li, Cao and Sun, Kun}]{Zhao2016}
Zhao L, Wu Y, Chen S, Ren Y, Chen P, Niu J, et~al.
\newblock {Feasibility Study on Prenatal Cardiac Screening Using
  Four-Dimensional Ultrasound with Spatiotemporal Image Correlation: A
  Multicenter Study}.
\newblock PLOS ONE 2016 6;11(6):e0157477.
\newblock \urlprefix\url{http://dx.plos.org/10.1371/journal.pone.0157477}.

\bibitem[{Lloyd et~al.(2018)Lloyd, David F A and Pushparajah, Kuberan and
  Simpson, John M and van Amerom, Joshua F P and Van Poppel, Milou and Schulz,
  Alexander and Kainz, Bernhard and Kuklisova-Murgasova, Maria and Lohezic,
  Maelene and Allsop, Joanna and Mathur, Sujeev and Bellsham-Revell, Hannah and
  {Trisha Vigneswaran} and Charakida, Marietta and Miller, Owen and Zidere,
  Vita and Sharland, Gurleen and Rutherford, Mary and Hajnal, Jo and Razavi,
  Reza}]{Lloyd2018}
Lloyd DFA, Pushparajah K, Simpson JM, van Amerom JFP, Van~Poppel M, Schulz A,
  et~al.
\newblock {High-resolution 3D visualisation of fetal congenital heart disease
  using prenatal MRI with motion corrected slice-volume registration (in
  press)}.
\newblock The Lancet 2018;.

\bibitem[{Keraudren et~al.(2015)Keraudren, Kevin and Kainz, Bernhard and Oktay,
  Ozan and Kyriakopoulou, Vanessa and Rutherford, Mary and Hajnal, Joseph V.
  and Rueckert, Daniel}]{KeraudrenMICCAI2015}
Keraudren K, Kainz B, Oktay O, Kyriakopoulou V, Rutherford M, Hajnal JV, et~al.
\newblock {Automated Localization of Fetal Organs in MRI Using Random Forests
  with Steerable Features}.
\newblock In: Medical Image Computing and Computer-Assisted Intervention –
  MICCAI, vol. 9351 Springer, Cham; 2015. p. 620--627.
\newblock
  \urlprefix\url{https://link.springer.com/chapter/10.1007/978-3-319-24574-4_74}.

\bibitem[{Demesmaeker et~al.(2017)Demesmaeker, Robin and Kober, Tobias and
  Yerly, Jérôme and Chaptinel, Jérôme and Prsa, Milan and Mivelaz, Yvan and
  Alamo, Leonor and Berchier, Gregoire and Rohner, Chantal and Stuber, Matthias
  and Piccini, Davide}]{Demesmaeker2017}
Demesmaeker R, Kober T, Yerly J, Chaptinel J, Prsa M, Mivelaz Y, et~al.
\newblock {Automated Heartbeat Detection for Self-Gated Fetal Cardiac MRI}.
\newblock In: International Society for Magnetic Resonance in Medicine; 2017.
  p. 0633.

\bibitem[{Kainz et~al.(2015)Kainz, Bernhard and Steinberger, Markus and Wein,
  Wolfgang and Murgasova, Maria and Malamateniou, Christina and Keraudren,
  Kevin and Aljabar, Paul and Rutherford, Mary and Hajnal, Joseph V. and
  Rueckert, Daniel and Kuklisova-Murgasova, Maria and Malamateniou, Christina
  and Keraudren, Kevin and Torsney-Weir, Thomas and Rutherford, Mary and
  Aljabar, Paul and Hajnal, Joseph V. and Rueckert, Daniel}]{Kainz2015}
Kainz B, Steinberger M, Wein W, Murgasova M, Malamateniou C, Keraudren K,
  et~al.
\newblock {Fast Volume Reconstruction from Motion Corrupted Stacks of 2D
  Slices.}
\newblock IEEE transactions on medical imaging 2015 3;34(9):1901--1913.
\newblock \urlprefix\url{http://ieeexplore.ieee.org/document/7064742/}.

\bibitem[{Kording et~al.(2018)Kording, Fabian and Yamamura, Jin and de Sousa,
  Manuela Tavares and Ruprecht, Christian and Hedstr{\"{o}}m, Erik and Aletras,
  Anthony H. and Ellen Grant, P. and Powell, Andrew J. and Fehrs, Kai and Adam,
  Gerhard and Kooijman, Hendrik and Schoennagel, Bjoern P.}]{Kording2018}
Kording F, Yamamura J, de~Sousa MT, Ruprecht C, Hedstr{\"{o}}m E, Aletras AH,
  et~al.
\newblock {Dynamic fetal cardiovascular magnetic resonance imaging using
  Doppler ultrasound gating}.
\newblock Journal of Cardiovascular Magnetic Resonance 2018 12;20(1):17.
\newblock
  \urlprefix\url{https://jcmr-online.biomedcentral.com/articles/10.1186/s12968-018-0440-4}.

\bibitem[{Larson et~al.(2004)Larson, Andrew C and White, Richard D and Laub,
  Gerhard and McVeigh, Elliot R and Li, Debiao and Simonetti, Orlando
  P}]{Larson2004}
Larson AC, White RD, Laub G, McVeigh ER, Li D, Simonetti OP.
\newblock {Self-gated cardiac cine MRI.}
\newblock Magnetic resonance in medicine 2004 1;51(1):93--102.
\newblock \urlprefix\url{https://dx.doi.org/10.1002%2Fmrm.10664}.

\bibitem[{Nijm et~al.(2008)Nijm, Grace M. and Sahakian, Alan V. and Swiryn,
  Steven and Carr, James C. and Sheehan, John J. and Larson, Andrew
  C.}]{Nijm2008}
Nijm GM, Sahakian AV, Swiryn S, Carr JC, Sheehan JJ, Larson AC.
\newblock {Comparison of self-gated cine MRI retrospective cardiac
  synchronization algorithms}.
\newblock Journal of Magnetic Resonance Imaging 2008 9;28(3):767--772.
\newblock \urlprefix\url{http://doi.wiley.com/10.1002/jmri.21514}.

\end{thebibliography}

\clearpage



\beginsupplement
\setcounter{page}{1}
\pagenumbering{roman}

\runningauthor{Supporting Information \ref{fig:simulation_results} \qquad Fetal Whole-Heart 4D MRI \qquad van Amerom et al.}
\null
\section*{Supporting Information \ref{fig:simulation_results}}

\setcounter{figure}{0}

\begin{figure}[htb]
\centering
\includegraphics[width = 1.0\textwidth]{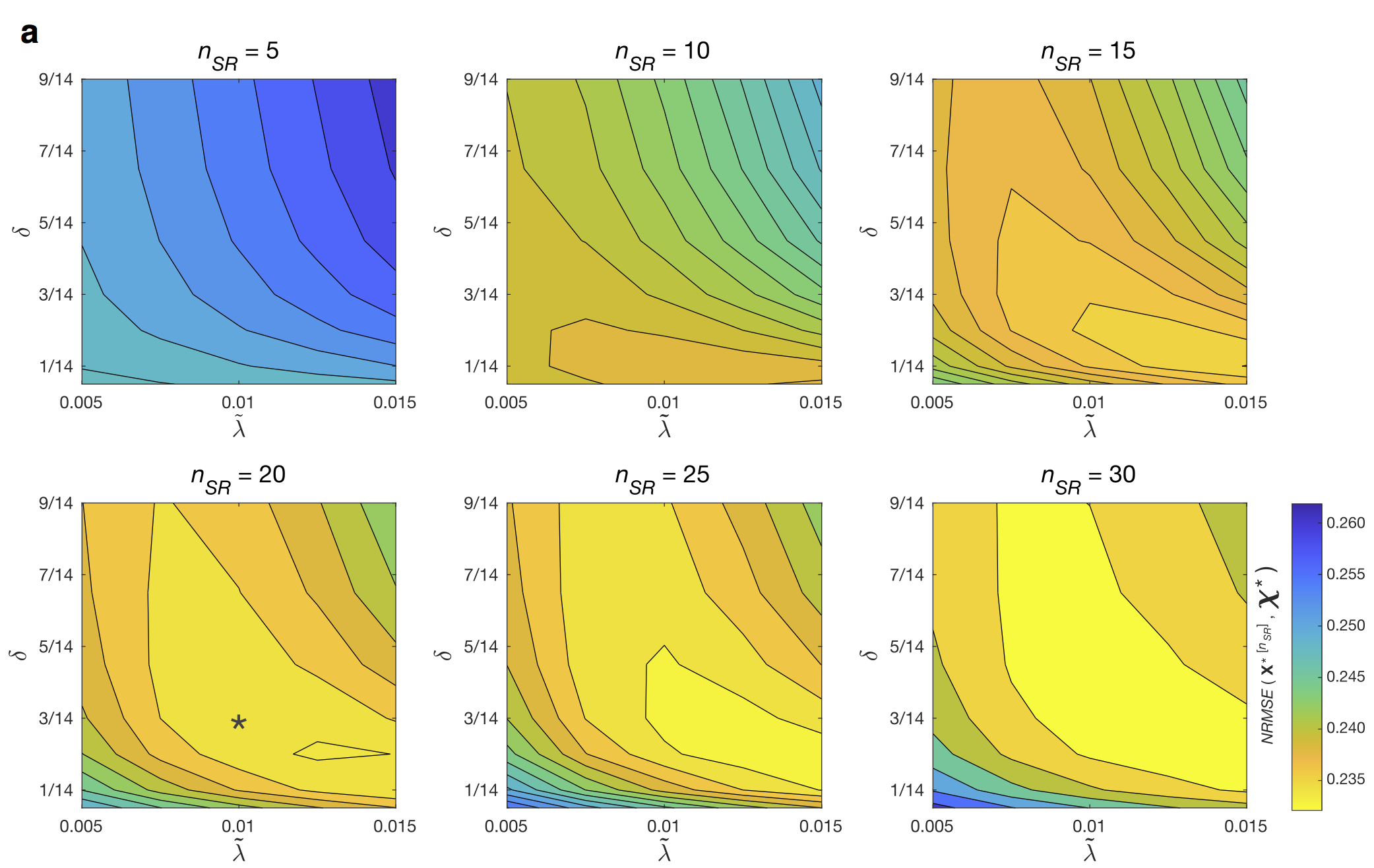} 
\caption[Reconstruction Parameter Optimisation Using Simulated MR Images]{(\textbf{a})~Normalised root mean square error ($NRMSE$) between 4D cine volume, $\mathbf{X}^*$, reconstructed from simulated MR images using known transformations and cardiac phases, and the ground truth cine volume, $\boldsymbol{\chi}$. 
Error contour maps are shown for $\mathbf{X}^{*\:[n_\text{SR}]}$ after $n_\text{SR}$~=~5 to~30 super-resolution iterations for a range of values for regularisation controlling parameter, $\widetilde{\lambda}$, and edge definition parameter,~$\delta$, used for edge-preserving regularisation. 
Values of $\delta$ are given relative to mean signal intensity. 
The star indicates the selected reconstruction parameter values.
}
\label{fig:simulation_nrmse_v_delta_v_lambda}
\end{figure}

\setcounter{figure}{0}

\begin{figure}[htb]
\centering
\includegraphics[width = 0.9\textwidth]{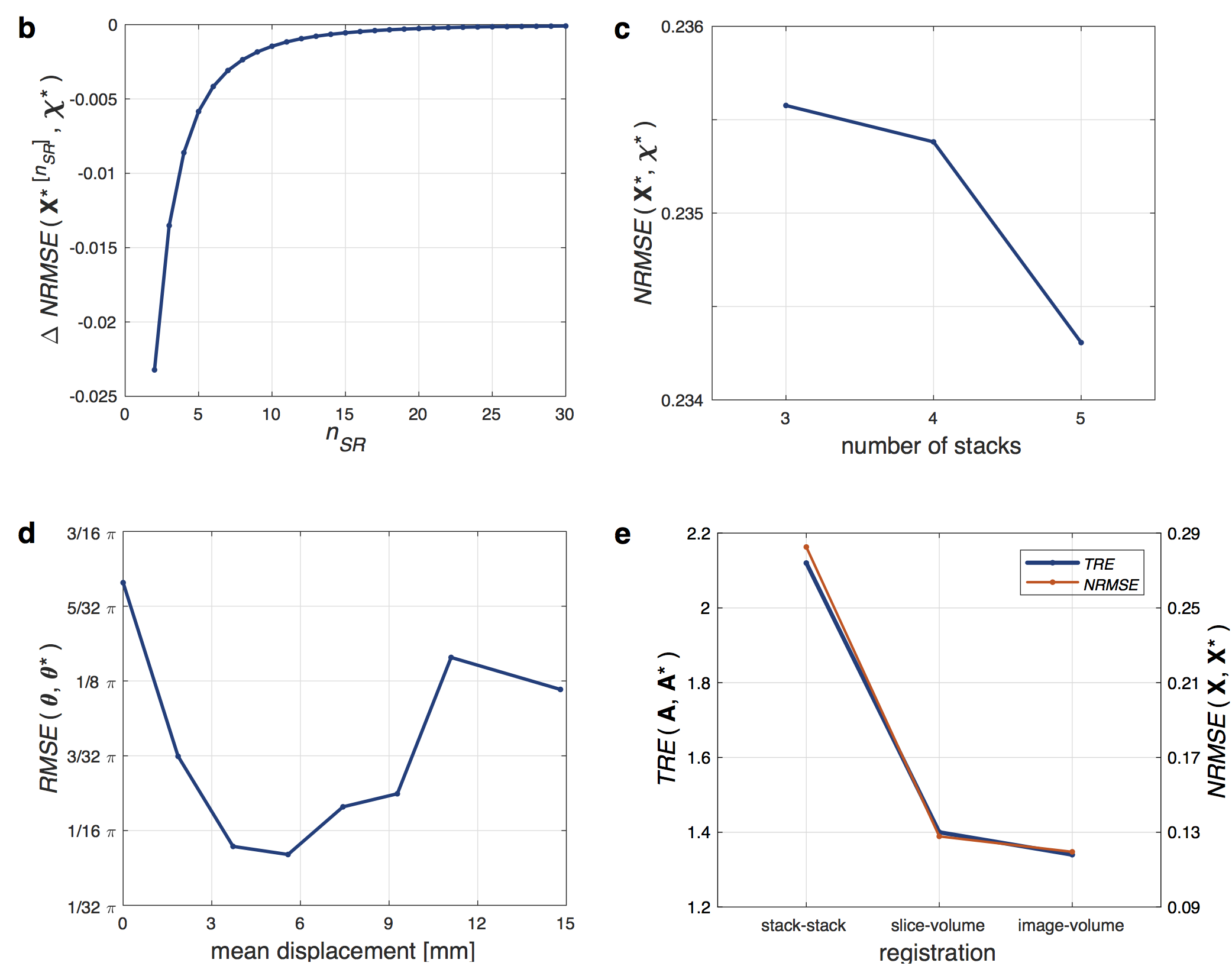}
\caption[Assessment of 4D Cine Reconstruction Using Simulated MR Images]{Assessment of 4D cine reconstruction using MR images simulated from numerical phantom. 
(\textbf{b})~Change in normalised root mean square error ($NRMSE$) between cine volume reconstructed using known transformations and cardiac phases, $\mathbf{X}^{*\:[n_\text{SR}]}$, and ground truth, $\boldsymbol{\chi}^*$, with number of number of super-resolution iterations, $n_\text{SR}$. 
Based on these results, $\widetilde{N}_\text{SR}$~=~20 iterations were used for the final volume reconstruction iteration when reconstructing fetal data. 
(\textbf{c})~$NRMSE$ versus number of stacks used in reconstruction showing a small decrease in error with increasing number of stacks, suggesting that all stacks available be used in the reconstruction. 
(\textbf{d})~Root mean square error~($RMSE$) of estimated cardiac phases after slice-slice cardiac synchronisation versus mean displacement of $\mathbf{A}^*$. The lowest errors occurred for ${\text{disp}(\mathbf{A}^*)}$ in the range of 2.3 to~9.3~mm with $RMSE(\boldsymbol{\theta},\boldsymbol{\theta}^*) < \frac{3}{32}\pi$, or approximately $0.05 \: t^{RR}$. 
The lowest $RMSE(\boldsymbol{\theta},\boldsymbol{\theta}^*)$ was $\frac{1}{20}\pi$, for simulated MR images with a ${\text{disp}(\mathbf{A}^*) = \text{5.6 mm,}}$, equivalent to 10~ms for the mean R-R interval measured in the fetal study~(404~ms). 
For reference, the median ${\text{disp}(\mathbf{A})}$ measured in Cohort~1 was 5.8~$\pm$~1.8~mm. 
Very low levels of movement resulted in reduced overlap between slices and, consequently, misalignment of the cardiac cycle in slices that had very little overlap with all other slices. 
Conversely, the overlap between slices increased with some movement, resulting in an improvement in cardiac synchronisation. 
However, large displacements lead to blurring in $\mathbf{X}_l$ and an increase in the cardiac synchronisation error for all slices. 
(\textbf{e})~The accuracy of motion correction was quantified as target registration error~(\textit{TRE}), defined as $TRE\left(\mathbf{A},\mathbf{A}^*\right)=\sum_{j}^{}\sum_{k}^{}\text{dist}\left(\mathrm{A}_k(y_{jk}),\mathrm{A}^*_k(y_{jk})\right)/\sum_{k}^{}N_{j}$, where $\text{dist}\left(\mathrm{A}_k(y_{jk}),\mathrm{A}^*_k(y_{jk})\right)$ is the spatial distance between the position of voxel $y_{jk}$ transformed by $\mathrm{A}_k$ and $\mathrm{A}^*_k$.
TRE is shown for estimated transformations for stack-stack, slice-volume, and frame-volume registrations~(blue line), with $NRMSE$ between 4D cine volumes reconstructed using estimated transformations and ground truth transformations using $N_\text{SR}$~=~20 iterations~(red line). 
TRE improved across the registration stages resulting in a final $TRE(\mathbf{A},\mathbf{A}^*)=\text{1.34 mm}$, equivalent to 2/3 the acquired in-plane resolution, after $N_\text{MC}$ = 3 frame-volume registration iterations, similar to the $TRE$ measured previously for fetal brain volume reconstruction 0.72~mm for 1~mm in-plane resolution images~\cite{Kuklisova-Murgasova2012}.
A similar decrease in $NRMSE$ was observed between cine volumes reconstructed with estimated transformations and ground truth transformations.
No clear improvement was observed for higher number of motion correction iterations, suggesting that $N_\text{MC} = \text{3}$ is sufficient.
}
\label{fig:simulation_results}
\end{figure}

\clearpage


\beginsupplement
\setcounter{page}{1}
\pagenumbering{roman}

\runningauthor{Supporting Information \ref{fig:id09_cine_vol_video} \qquad Fetal Whole-Heart 4D MRI \qquad van Amerom et al.}
\null
\section*{Supporting Information \ref{fig:id09_cine_vol_video}}

\setcounter{figure}{1}

\begin{figure}[htb]
\centering
\bigskip
\includegraphics[width = 0.42\textwidth]{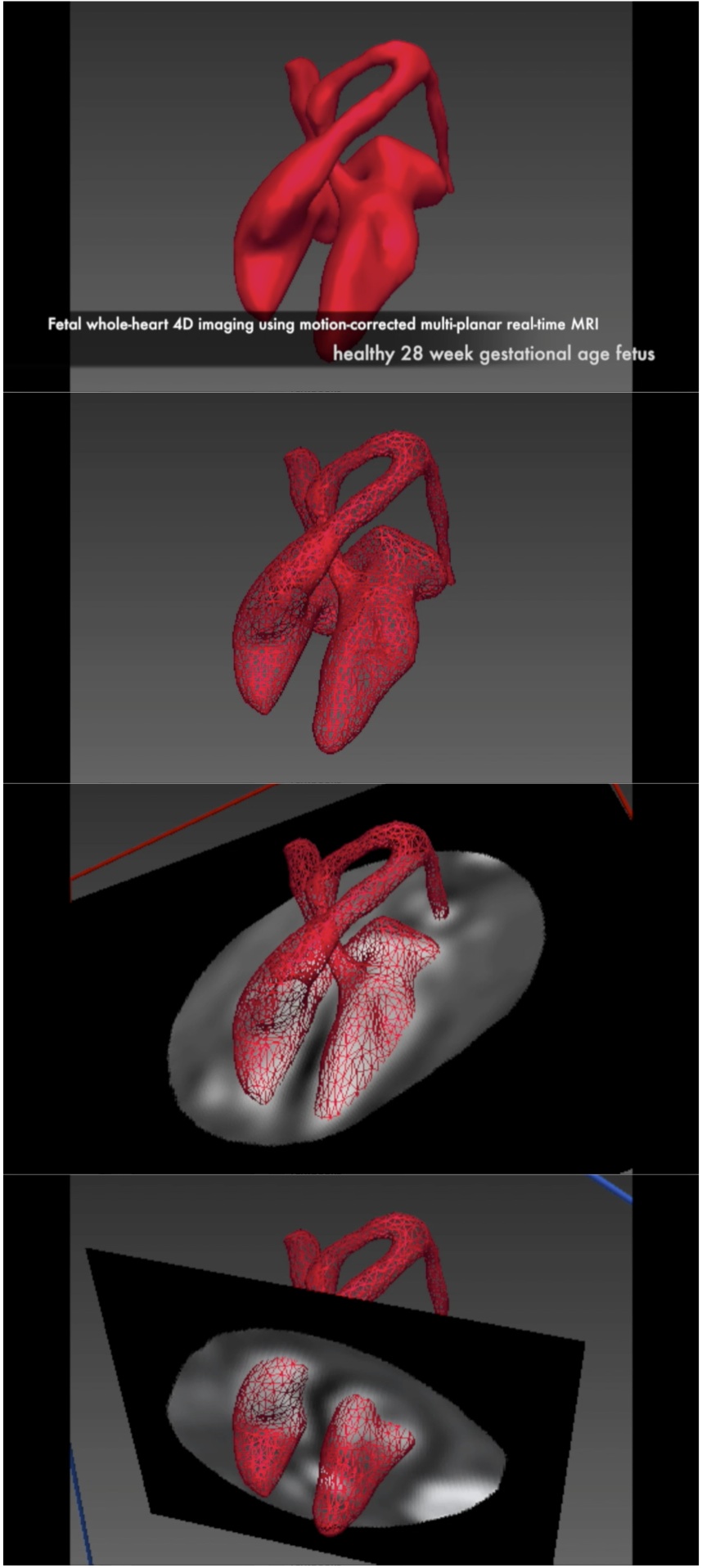}
\caption[Animation Demonstrating Spatial and Temporal Aspects of 4D Whole-Heart Reconstruction]{Animation demonstrating spatial and temporal aspects of 4D whole-heart reconstruction of healthy 28\textsuperscript{+0}~week gestational age fetus~(ID09) shown in Figure~\ref{fig:id09_cine_vol_views}, with volume rendering of blood pool (red) for reference.
Video can be viewed at \href{http://doi.org/10.6084/m9.figshare.7413566}{http://doi.org/10.6084/m9.figshare.7413566}.}
\label{fig:id09_cine_vol_video}
\end{figure}

\clearpage

\beginsupplement
\setcounter{page}{1}
\pagenumbering{roman}

\runningauthor{Supporting Information \ref{fig:id09_rr_v_time} \qquad Fetal Whole-Heart 4D MRI \qquad van Amerom et al.}
\null
\section*{Supporting Information \ref{fig:id09_rr_v_time}}

\setcounter{figure}{2}

\begin{figure}[htb]
\centering
\includegraphics[width = 1.0\textwidth]{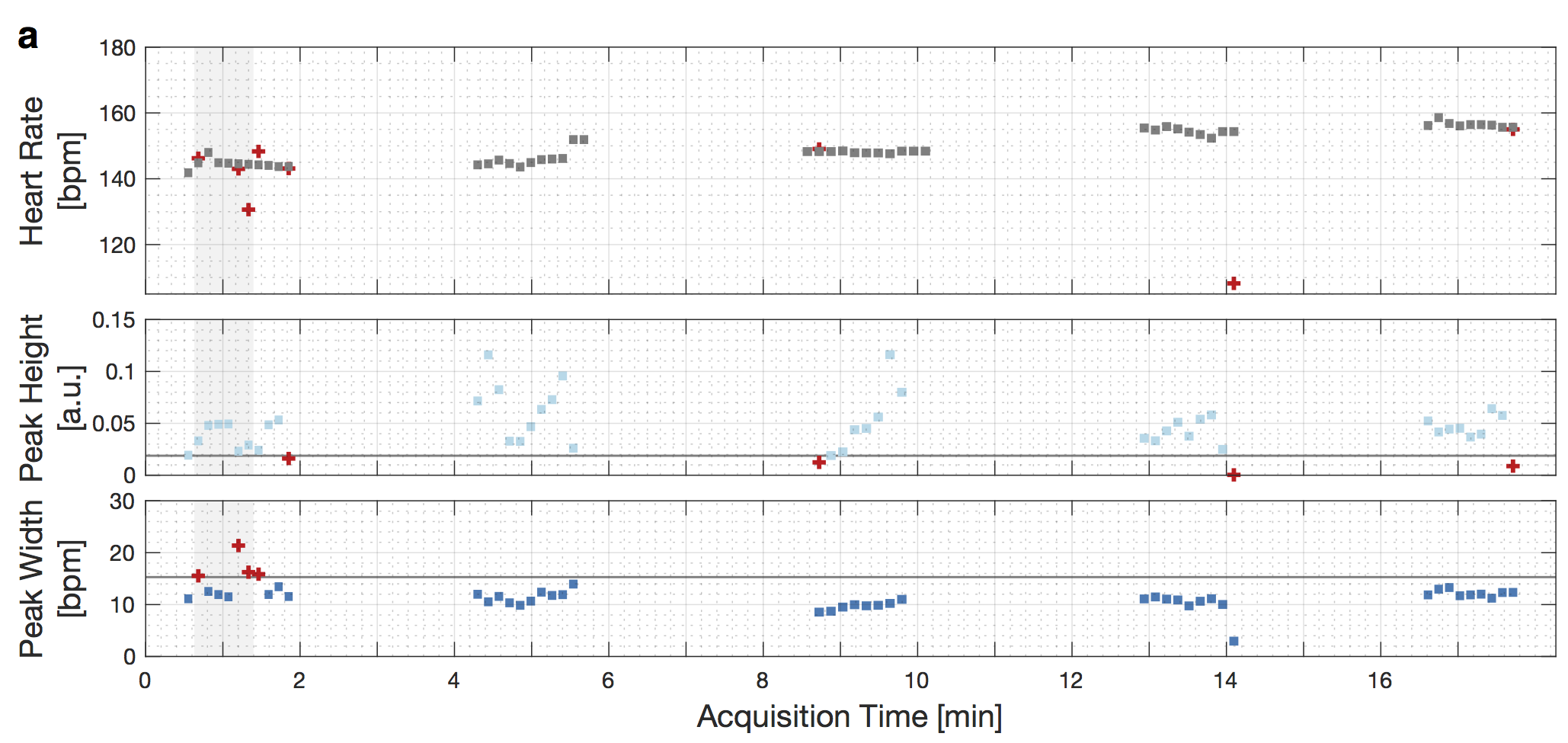}
\caption[Estimated Heart Rates in a Healthy Fetus]{(\textbf{a})~Estimated heart rate~(top row) for each slice acquired in a healthy 28\textsuperscript{+0}~week gestational age fetus~(ID09), with mean heart rate 150$\pm$5~bpm~(401$\pm$13~ms). Unreliable heart rate estimates (red crosses) were identified using the height~(middle row) and width~(bottom row) of the peak in the temporal frequency spectrum used to estimate the heart rate, as shown in Fig. \ref{fig:method_cardiac_synchronisation}. Threshold limits~(horizontal lines, peak height~0.02, peak width~15.5~bpm) were calculated as three scaled median absolute deviations from the median. The heart rates for slices with peak heights less than the threshold were replaced with values linearly interpolated from temporally adjacent slices. Subsequently, heart rates for slices with peak widths above the threshold were replaced in a similar manner. Motion correction and outlier rejection results are shown in Fig.~\ref{fig:id09_transformations_and_probability_v_time}b for the all image frames acquired in the time window indicated by the grey vertical band~(38 to 83~seconds).}
\label{fig:id09_rr_v_time}
\end{figure}

\setcounter{figure}{2}

\begin{figure}[htb]
\centering
\includegraphics[width = 1.0\textwidth]{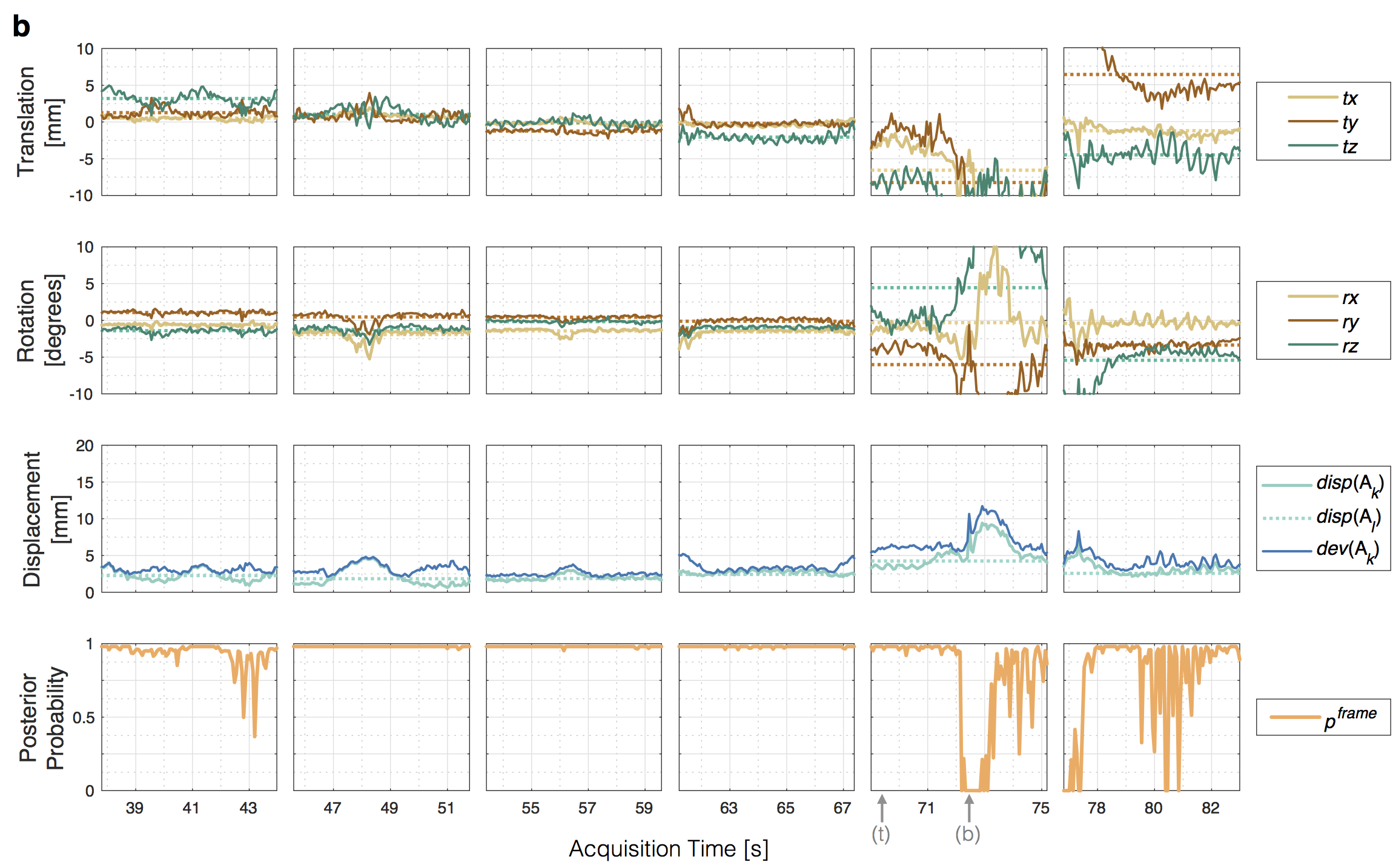}
\caption[Estimated Motion Correction Transformations and Image Frame-Wise Probabilities in a Healthy Fetus]{(\textbf{b}) Estimated transformations, displacements and image frame-wise probabilities for six consecutive slices in one stack of multi-planar dynamic MR images acquired in a 28\textsuperscript{+0}~week gestational age fetus~(ID09), corresponding to time window marked in Fig.~\ref{fig:id09_rr_v_time}a. 
An episode of fetal movement can be seen from acquisition time 72-78~s in the fifth and sixth slices shown.
Translations \textit{tx}, \textit{ty} and \textit{tz} are with respect to scanner right-left, anterior-posterior and superior-inferior directions, respectively. Rotations \textit{rx}, \textit{ry} and \textit{rz} are about scanner y-z, x-z and x-y axes, respectively. Translation and rotation of the average slice transformation are plotted as dotted lines. Displacement of image frame-wise transformations, $dev(\mathrm{A}_k) = \sum_{j}^{} dist\left(\mathrm{A}_l(y_{jk}),\mathrm{A}_k(y_{jk})\right) / N_{j}$, and average slice transformations, $\text{disp}(\mathrm{A}_l) = \sum_{k \in \text{slice} l}\sum_{j}^{} dist\left(\mathrm{A}_l(y_{jk}),\mathrm{A}_k(y_{jk})\right) / \sum_{k \in \text{slice} l} N_{j}$, are plotted as solid and dotted lines, respectively. Deviation from the average slice transformation, $\text{dev}(\mathrm{A}_k)$, is shown relative to $\text{disp}(\mathrm{A}_l)$. In the slices shown, $\text{dev}(\mathbf{A}_l)$~=~0.8, 1.4, 0.6, 1.0, 2.8 and 1.6~mm. 
Arrows on the time axis of the fifth slice indicate the image frames shown on the top (t) and bottom (b) rows in  Fig.~\ref{fig:id09_outlier_rejection_images}c.}
\label{fig:id09_transformations_and_probability_v_time}
\end{figure}

\setcounter{figure}{2}

\begin{figure}[htb]
\centering
\includegraphics[width = 0.57\textwidth]{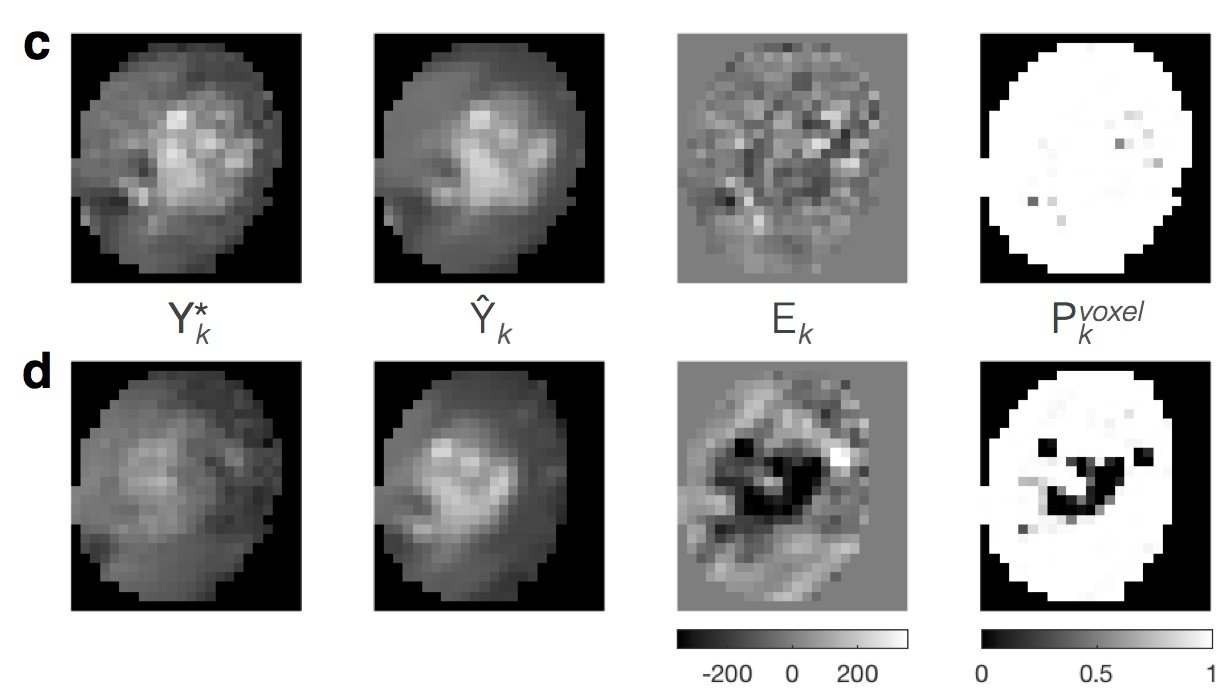}
\caption[Outlier Rejection in the Presence of Motion]{Voxel-wise error and probability maps for two image frames from the same slice acquired in high transverse plane in a 28\textsuperscript{+0}~week gestational age fetus~(ID09) at completion of 4D cine reconstruction. Image frames correspond to markers in Fig.~\ref{fig:id09_transformations_and_probability_v_time}b, for (\textbf{c})~frame acquired during a period of no fetal movement and (\textbf{d})~frame acquired during a period of fetal movement (bottom row). Cropped views of the fetal heart show, from left to right,  intensity-corrected images~($\mathrm{Y}^*_k$), images~($\widehat{\mathrm{Y}}_k$) generated using Eq.~\ref{eq:mr_image_aquisition_model}, error maps~($\mathrm{E}_k$) and voxel-wise probability maps~($\mathrm{P}_k^\text{voxel}$).}
\label{fig:id09_outlier_rejection_images}
\end{figure}

\setcounter{figure}{2}

\begin{figure}[htb]
\centering
\includegraphics[width = 0.74\textwidth]{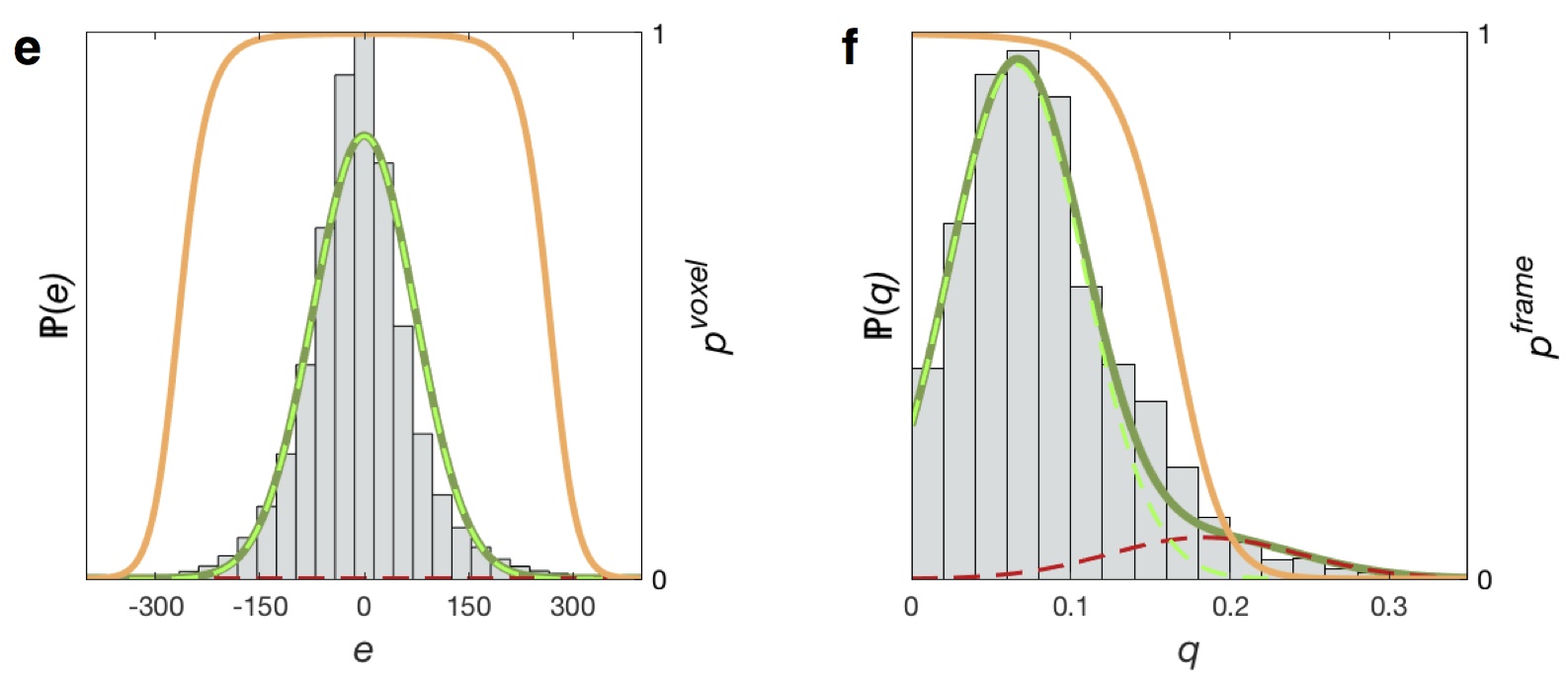}
\caption[Outlier Rejection Using Robust Statistics in a Healthy Fetus]{Outlier rejection using voxel- and image frame-wise robust statistics in a 28\textsuperscript{+0}~week gestational age fetus~(ID09) at completion of 3D cine reconstruction. (\textbf{e})~Voxel-wise error distribution with likelihood, $\mathbb{P}(e)$ (solid green line), of observing error, $e$, modelled as the mixture of a Gaussian-distributed inlier class (dashed green line) and uniformly-distributed outlier class (dashed red line). Distribution parameters were estimated by maximising the log-likelihood of $\mathbb{P}(e)$ and used to map error to voxel-wise probability, $p^\text{voxel}$ (solid orange line). (\textbf{f})~Distribution of image frame potentials, $q$, with likelihood, $\mathbb{P}(q)$ (solid green line), modelled as the mixture of Gaussian-distributed inlier (dashed green line) and outlier (dashed red line) classes, with expectation maximisation of $\mathbb{P}(q)$ resulting in frame-wise probability weighting, $p^\text{frame}$ (solid orange line).}
\label{fig:id09_outlier_rejection_distribution}
\end{figure}

\clearpage


\beginsupplement
\setcounter{page}{1}
\pagenumbering{roman}

\runningauthor{Supporting Information \ref{fig:id04_realtime_and_cine_images} \qquad Fetal Whole-Heart 4D MRI \qquad van Amerom et al.}
\null
\section*{Supporting Information \ref{fig:id04_realtime_and_cine_images}}

\setcounter{figure}{3}

\begin{figure}[!hb]
\centering
\includegraphics[width = 0.95\textwidth]{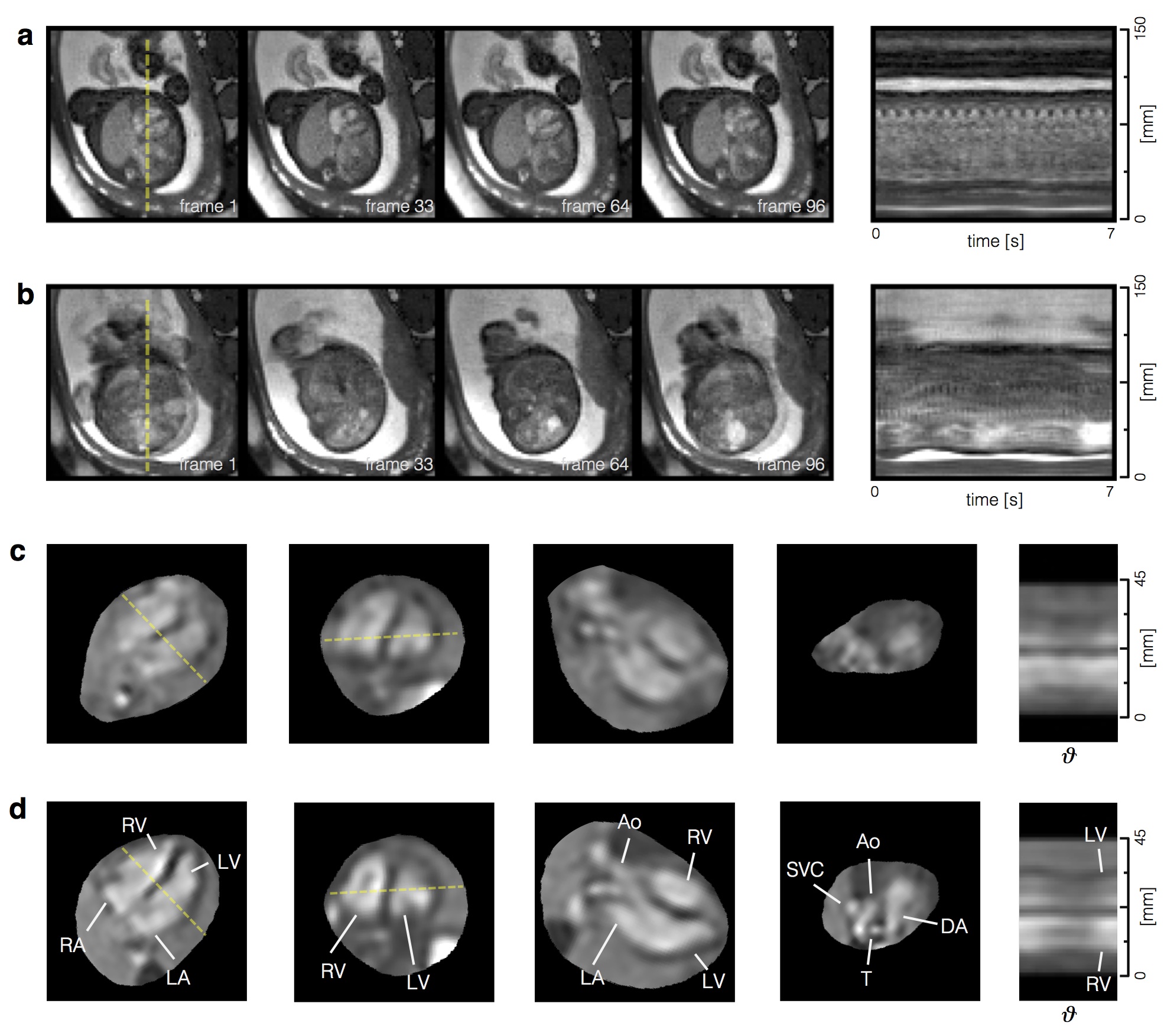}
\caption[Impact of Fetal Movement on Acquired Dynamic Images and Reconstructed 4D Volume]{Dynamic MR images and reconstructed 4D heart of a 29\textsuperscript{+6} week gestational age fetus~(ID04) with isolated right aortic arch. 
Cropped views of selected image frames in one slice and line profile across the fetal heart and chest corresponding to yellow dashed line for all frames in the slice for (\textbf{a})~a slice with little fetal movement, ${\text{dev}(\mathbf{A}_l) = \text{0.9 mm}}$, and (\textbf{b})~a slice with large fetal movements, ${\text{dev}(\mathbf{A}_l) = \text{10.1 mm}}$, from the same stack. 
(\textbf{c})~Reconstruction using all acquired dynamic images resulting in a 4D volume corrupted by motion. 
(\textbf{d})~Exclusion of~19 of~54 total slices resulting in a 4D reconstruction with improved quality, particularly in the definition and detail of the arch anatomy in question.
The heart is shown in (c) and (d) at end-ventricular diastole in, from left to right: four chamber view, short axis view, three chamber view, high transverse view, similar to a three vessel view, and as a line profile at the intersection of the long and short axes (dashed yellow lines) showing a cross section of the ventricles across cardiac phases ($\vartheta$). 
The aortic~(Ao) arch can be seen emerging from the left ventricle~(LV), and then passing between the left atrium~(LA) and right ventricle~(RV) in the three vessel view.
The vascular ring can be also be seen in the high transverse view, with the superior vena cava~(SVC) at right and the duct arch~(AD) passing to the left of the trachea~(T), as normal, while the aorta~(Ao) passes to the right. 
Manual exclusion of data was required in two of twenty cases~(ID04,ID07).}
\label{fig:id04_realtime_and_cine_images}
\end{figure}

\clearpage

\end{document}